\documentclass[preprint,showpacs,preprintnumbers,amsmath,amssymb,aps]{revtex4}
\usepackage{graphicx}
\usepackage{dcolumn}
\usepackage{bm}
\usepackage[dvips]{color}
\begin{document}

\title{Piezo-control of magnetic anisotropy in GaMnAs: \\
Reversible manipulation of magnetization orientation \\
and irreversible magnetization switching}
\author{C. Bihler\footnote{Christoph.Bihler@wsi.tum.de}$^1$, M. Althammer\footnote{Matthias.Althammer@wmi.badw-muenchen.de}$^2$, A. Brandlmaier$^2$, S. Gepr\"{a}gs$^2$, M. Weiler$^2$, M. Opel$^2$, W. Schoch$^3$, W. Limmer$^3$, R. Gross$^2$, M. S. Brandt$^1$, and S.~T.~B. Goennenwein\footnote{Sebastian.Goennenwein@wmi.badw-muenchen.de}$^2$}
\affiliation{$^1$Walter Schottky Institut, Technische Universit\"{a}t M\"{u}nchen, Am
Coulombwall 3, 85748 Garching, Germany}
\affiliation{$^2$Walther-Meissner-Institut, Bayerische Akademie der Wissenschaften,
Walther-Meissner-Str.~8, 85748 Garching, Germany}
\affiliation{$^3$Institut f\"{u}r Halbleiterphysik, Universit\"{a}t Ulm, 89069 Ulm, Germany}

\begin{abstract}
We have investigated the magnetic properties of a piezoelectric actuator/ferromagnetic semiconductor hybrid structure.
Using a GaMnAs epilayer as the ferromagnetic semiconductor and applying the piezo-stress along its [110] direction, we quantify the magnetic anisotropy as a function of the voltage $V_{\mathrm{p}}$ applied to the piezoelectric actuator using anisotropic magnetoresistance techniques. We find that the easy axis of the strain-induced uniaxial magnetic anisotropy contribution can be inverted from the [110] to the $\left[1\overline{1}0\right]$ direction via the application of appropriate voltages $V_{\mathrm{p}}$. At $T=5$~K the magnetoelastic term is a minor contribution to the magnetic anisotropy. Nevertheless, we show that the switching fields of $\rho(\mu_0H)$ loops are shifted as a function of $V_{\mathrm{p}}$ at this temperature. At 50~K $-$ where the magnetoelastic term dominates the magnetic anisotropy $-$ we are able to tune the magnetization orientation by about $70^\circ$ solely by means of the electrical voltage $V_{\mathrm{p}}$ applied. Furthermore, we derive the magnetostrictive constant $\lambda_{111}$ as a function of temperature and find values consistent with earlier results. We argue that the piezo-voltage control of magnetization orientation is directly transferable to other ferromagnetic/piezoelectric hybrid structures, paving the way to innovative multifunctional device concepts. As an example, we demonstrate piezo-voltage induced irreversible magnetization switching at $T=40$~K, which constitutes the basic principle of a nonvolatile memory element.
\end{abstract}

\maketitle

\section{\label{sec:introduction}INTRODUCTION}

Ferromagnetic semiconductors unite the long-range magnetic ordering characteristic of ferromagnets with the versatile properties of conventional semiconductors. This class of multifunctional materials therefore is very attractive from a fundamental physics point of view. Moreover, novel spin-electronic devices can be realized using ferromagnetic semiconductors, in which the electronic functionality is directly linked to magnetic properties such as the magnetization orientation \cite{Ohn98, Die00, Wol01, Rue03, Pea05, Jun2006, Fig07, Pap07}. For device applications it is attractive to control the magnetic properties via a \textit{non-magnetic} parameter. In this paper, we investigate the approach to strain a dilute magnetic semiconductor (DMS) lattice by means of an external piezoelectric actuator in order to manipulate its magnetic properties. Using GaMnAs as the DMS, we discuss in detail the piezo-voltage control of the magnetic anisotropy, which can be utilized to reversibly manipulate magnetization orientation as well as to irreversibly switch the magnetization between two magnetic easy axis orientations. The experiments discussed here pave the way for the three dimensional control of spins via a non-magnetic control parameter.

We begin with a short introduction into GaMnAs, before reviewing different approaches to influence the magnetic properties of magnetic semiconductors via non-magnetic parameters reported in literature. Ga$_{1-x}$Mn$_x$As is the prototype ferromagnetic semiconductor. Although its Curie temperature $T_{\rm C} < 170$~K is still below room temperature, it has been investigated vigorously in the last decade \cite{Jun2006}. The effect of the Mn incorporation is twofold: On the one hand, the half-filled $d$-shells of the Mn$^{2+}$ ions provide localized magnetic moments with spin $S=5/2$. On the other hand, Mn also is a shallow acceptor in GaAs, so that ferromagnetic GaMnAs doped with a few percent of Mn is degenerately p-type. These itinerant holes mediate the ferromagnetic exchange between the localized Mn moments \cite{Die00, Die01}, linking the magnetic properties of GaMnAs to the GaAs valence band structure \cite{Die01}. The strong $p$-$d$ coupling between the delocalized holes and the localized electrons in the Mn $d$-shell results in large magnetoresistive effects in GaMnAs. For example, the anisotropic magnetoresistance (AMR) has a "giant" transverse (planar Hall) component \cite{Tan03}. The anomalous Hall effect due to the magnetization in GaMnAs is routinely exploited for magnetic characterization purposes \cite{Ohn01}. Moreover, the spin polarization of the itinerant holes is large, with values of up to $85\%$ reported in literature \cite{Dor04, Bra03, Tan01}. 
Since the magnetic domains extend over 100 microns or more \cite{Wel03}, GaMnAs is ideally suited as a spin-injecting or spin-polarizing contact in spintronic devices. Indeed, novel functionalities or physical effects have already been demonstrated in spin-electronic devices based on GaMnAs, such as the emission of circularly polarized light from a so-called spin light emitting diode \cite{Dor04, Ohn99}, the controlled motion of domain walls via current pulses \cite{Yam04, Yam06}, the tunneling anisotropic magnetoresistance in GaMnAs/insulator/normal metal structures \cite{Gou04}, and non-volatile memory device concepts \cite{Rue03, Fig07, Pap07}.

Ga$_{1-x}$Mn$_x$As layers grown on GaAs are compressively strained, leading to a first-order uniaxial magnetic anisotropy so that the film plane is an easy magnetic plane for the typical range of Mn concentrations ($0.01<x<0.1$) \cite{Liu03, Liu2005, She97}. A smaller cubic (biaxial) anisotropy yields two approximately orthogonal easy axes within the film plane. Furthermore, a small in-plane uniaxial anisotropy along $\left[110\right]$ is present in most GaMnAs films on 
GaAs \cite{Saw04, Ham03, Wel04, Saw05, Sta05, Ham05, Wan05, Ham06a, Ham06b}, the microscopic origin of which is still controversially discussed. In GaMnAs films with tensile strain, in contrast, the film plane is magnetically hard \cite{Liu03, She97}, so that the magnetization aligns along the film normal at low magnetic fields. Furthermore, the magnetic anisotropy in GaMnAs qualitatively changes with the temperature. In comparison to other ferromagnetic metals, demagnetization effects only play a minor role in GaMnAs, as the saturation magnetization is small due to the small concentration of magnetic ions.

One of the most intriguing properties of DMS is the strong dependence of their magnetic properties on non-magnetic parameters, such as electric field \cite{Ohn00, Chi03a}, light irradiation \cite{Kos97, Bou02, Liu04}, temperature \cite{Ham03, Saw05, Mas05}, dopant density \cite{Tit05,Tak02}, strain \cite{She97, Liu03, Liu2005, Dae07}, or pressure \cite{Cso05}. This allows for various control schemes of the magnetic properties of DMS, which we shortly discuss in the following.

Ohno \textit{et al.} \cite{Ohn00} demonstrated that an electric field-control of ferromagnetism is possible in InMnAs samples covered with a metallic gate. Upon the application of an appropriate voltage to the gate electrode, the density of itinerant holes is altered, which allows to switch on or off the ferromagnetic exchange. Electrically assisted magnetization reversal, as well as electrical demagnetization have already been demonstrated in devices utilizing this effect \cite{Chi03a}. While this electric field-control of ferromagnetism appears very attractive for spintronic devices, the high carrier density in degeneratly doped DMS and the resulting screening effects limit the thickness of magnetic semiconductor films controllable by electric fields to a few nm at most, imposing severe constraints on the device geometries.

Another approach to vary the hole concentration and thus the magnetic properties relies on the introduction of additional non-magnetic dopant species into GaMnAs. By co-doping GaMnAs with Be, the relative strength of the cubic and uniaxial magnetic anisotropy contributions can be altered \cite{Tit05}. The incorporation of up to $17\%$ of Al into GaMnAs significantly reduces the density of free holes, eventually leading to a magnetically hard film plane \cite{Tak02}. The incorporation of hydrogen also allows to control the hole density in GaMnAs \cite{Goe04}. The hydrogen atom forms a complex with the Mn ion \cite{Bou03, Bra04} and thereby passivates its acceptor function, while the localized 3\textit{d} electrons stay unaffected. The hydrogen incorporation is stable at ambient temperatures, but can be reversed by a mild thermal as well as a pulsed-laser annealing \cite{The05, Far08}. However, while the exchange coupling in GaMnAs films with thicknesses of several hundred nm can be switched on or off via co-doping or hydrogenation, these techniques do not allow to manipulate the magnetization orientation while operating a spintronic device.

For the realization of DMS tunable during device operation, the dependence of the magnetic properties on the crystal lattice appears much more promising. Csontos \textit{et al.} showed that the ferromagnetic exchange interaction in InMnSb can be controlled via the application of hydrostatic pressure \cite{Cso05}. As discussed above, the magnetic anisotropy in GaMnAs was found to qualitatively change as a function of strain \cite{She97, Liu03, Liu2005, Dae07}. Recently also the control of the mangnetic anisotropy via anisotropic strain relaxation in patterened structures has been reported \cite{Wen07}. These experimental observations are explained by the established theoretical model for the ferromagnetic exchange \cite{Die01}: The strain alters the symmetry of the hole wave function in the GaAs valence band, which due to the hole-mediation of ferromagnetic exchange has a strong influence on the magnetic anisotropy. This picture also is compatible with the recent measurements of the magneto-elastic constants of GaMnAs in micro-mechanical devices \cite{Mas05}.

In the following we will investigate the control of the magnetic anisotropy of GaMnAs via the application of an external stress by a piezoelectric actuator. This approach has already been shown to allow in-situ manipulation of magnetic anisotropy in piezoelectric actuator/ferromagnet hybrid samples at room temperature \cite{Bot06, Bra08}. We have recently demonstrated that this approach can be transfered to piezoelectric actuator/GaMnAs hybrids \cite{Goe08}, as also reported by \cite{Rus08, Ove08}. Here we compile our extensive experimental findings and show that they can be understood consistently using a free energy model. In particular, we make use of the strong temperature dependence of the magnetic anisotropy in GaMnAs to tune the free energy surface so that the effect of strain on the magnetic properties is maximal. 

This paper is organized as follows: After a review of the magnetic free energy approach in Sec.~\ref{sec:F} we summarize the experimental procedures employed and the properties of the piezoelectric actuator used in Sec.~\ref{sec:experimental} and \ref{sec:strain}, respectively. The method to determine the magnetic anisotropy from magnetotransport experiments is described in Sec.~\ref{sec:limmer}. At low temperatures ($T\approx5$~K) the piezo-voltage induced changes are small compared to the dominating cubic contribution to magnetic anisotropy in GaMnAs. In this temperature range we demonstrate the piezo-voltage control of the magnetoresistance switching fields in external magnetic field sweeps (Sec.~\ref{sec:switching5K}). In contrast, at $T\approx50$~K the magnetoelastic contribution dominates magnetic anisotropy (Sec.~\ref{sec:50K}), which allows to continuously and reversibly rotate magnetization orientation (Sec.~\ref{sec:rotation}). From a comparison of the piezo-induced strain (Sec.~\ref{sec:strain}) and the corresponding magnetoelastic contribution to magnetic anisotropy we are able to derive the temperature dependence of the magnetostrictive constant (Sec.~\ref{sec:lambda}). Finally, we demonstrate a piezo-voltage induced irreversible magnetization switching (Sec.~\ref{sec:NVswitching}) and discuss possible applications of the investigated effects (Sec.~\ref{sec:conclusion}).

\section{\label{sec:F}FREE ENERGY APPROACH}

\begin{figure}[tbp]
\includegraphics[width=0.65\textwidth]{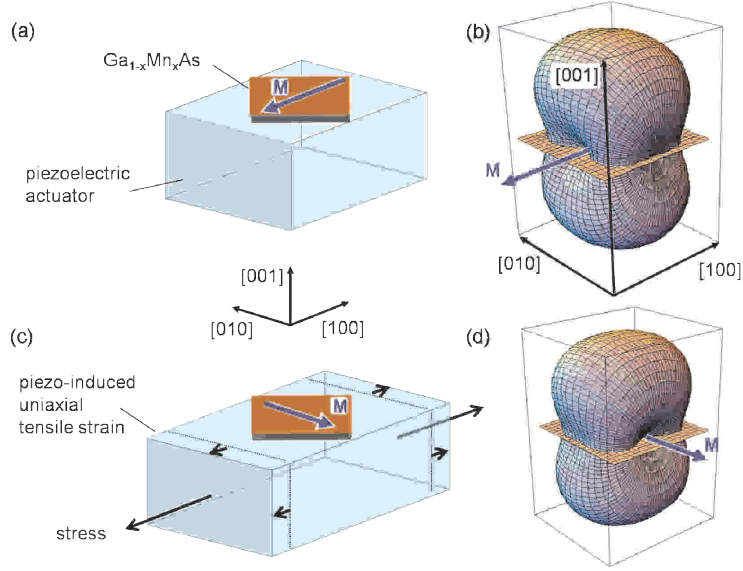} 
\caption{(color online) (a) Schematic illustration of a 90$^\circ$ magnetization switching for a GaMnAs film affixed onto a piezoelectric actuator with the magnetic easy axis $\left[100\right]$ parallel to the main expansion direction of the actuator. (b) Corresponding free energy surface. The free energy for the magnetization oriented along a specific direction is given by the distance from the center of the coordinate system to the surface. In this example the magnetization is oriented along $\left[\overline{1}00\right]$. (c) A positive voltage applied to the actuator leads to an elongation of the actuator/GaMnAs layer. (d) This strain induces a uniaxial anisotropy with a magnetic hard axis in elongation direction, which leads to a change in the relative strength of the in-plane magnetic easy axes [Fig.~\ref{fig:schema2}(a)] and thus a 90$^\circ$ magnetization switching into the global minimum at $\left[0\overline{1}0\right]$.}
\label{fig:schema}
\end{figure}

\begin{figure}[tbp]
\includegraphics[width=0.7\textwidth]{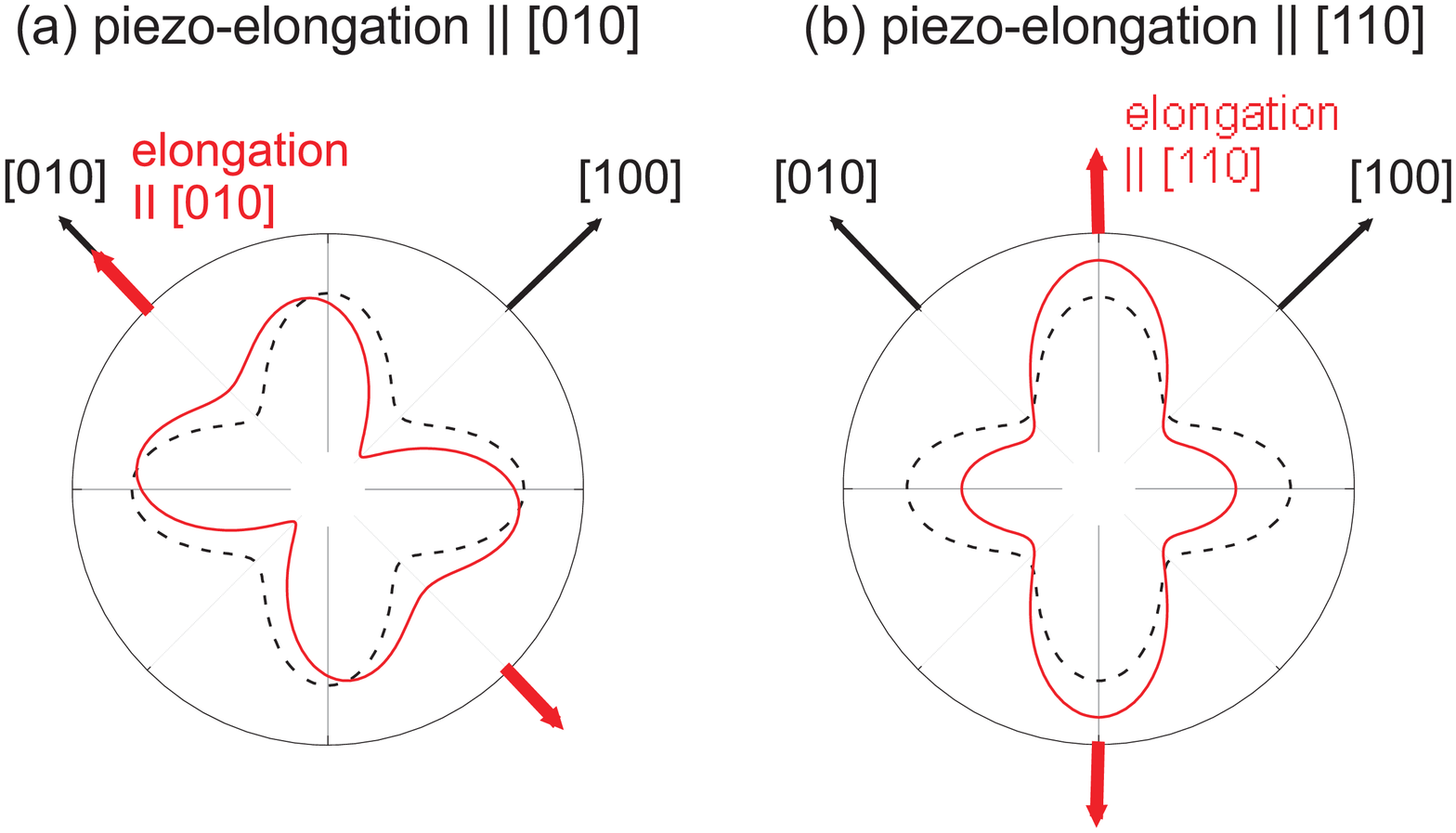} \vspace{-2.0cm}
\caption{(color online) Cuts through the free energy surface within the film plane for two different strain configurations. The black dashed curves illustrate the in-plane cubic anisotropy contribution with magnetic easy axes along $\left<100\right>$. (a) For the main elongation direction of the piezoelectric actuator aligned along a magnetic easy axis the strain-induced uniaxial anisotropy leads to a change of the relative strengths of the magnetic easy axes (red full curve). (b) In contrast, for the main elongation direction of the piezoelectric actuator aligned along a locally hard axis of the in-plane cubic anisotropy, \textit{e.g.} $\left[110\right]$, the strain-induced uniaxial anisotropy leads to a rotation and a change of the relative orientations of the magnetic easy axes (red full curve).}
\label{fig:schema2}
\end{figure}

To describe the magnetic anisotropy and thus magnetization orientation in our sample we adopt a free energy approach. To second order, the free energy in GaMnAs normalized to the saturation magnetization is given by \cite{Liu2005, Lim06}
\begin{eqnarray}
F/M&=&-\mu_0H(\textbf{h}\cdot\textbf{m})+B_{\rm c||}(m^{4}_{x}+m^{4}_{y})+B_{110}(V_{\rm p})(\textbf{j}\cdot\textbf{m})^2+B_{010}m^{2}_{y}= \nonumber \\ 
&=& -\mu_0H \cos(\beta-\alpha)+B_{\rm c||}\left[ \cos^4(\beta+45^\circ) + \cos^4(\beta-45^\circ)\right] + \nonumber \\
& &    + B_{110}(V_{\rm p})\cos^2\beta+B_{010}\cos^2(\beta-45^\circ),
\label{eq:F}
\end{eqnarray}
where we have assumed that the magnetization is aligned within the film plane. $\mu_0=4\pi\times10^{-7}~\frac{\rm Vs}{\rm Am}$ is the vacuum induction, $m_i$ are the direction cosines of the magnetization relative to the cubic axes,   $B_{\rm c||}$ is the in-plane cubic anisotropy parameter, and $B_{110}(V_{\rm p})$ and $B_{010}$ are the uniaxial anisotropy parameters. The orientation of the external magnetic field $H$ and the magnetization $M$ are given by the unit vectors \textbf{h} and \textbf{m}, with the corresponding angles $\alpha$ and $\beta$ with respect to the current direction \textbf{j} as shown in Fig.~\ref{fig:Experimental}. The individual terms in (\ref{eq:F}) describe the Zeeman interaction, the in-plane cubic and two uniaxial magnetic anisotropy contributions along [110] and [010] crystallographic directions. The third term $\left[B_{110}(V_{\rm p})(\textbf{j}\cdot\textbf{m})^2\right]$, the magnetoelastic contribution to the magnetic anisotropy, accounts for the changes in magnetic anisotropy due to a voltage $V_{\rm p}$ applied to the piezoelectric actuator in the particular stress geometry discussed below. The fact that the piezo-voltage induced effect can be described by a uniaxial anisotropy is derived in the Appendix. Note that the anisotropy fields used in Ref.~\cite{Bihler3112006,Bih07} are given by $\frac{2K^{||}_{\rm c1}}{M}=-4B_{\rm c||}$ for the in-plane cubic field and $\frac{2K^i_{\rm u1}}{M}=2B_{i}$ for the uniaxial fields.

Figure~\ref{fig:schema}(b) illustrates a typical free energy surface of a compressively strained GaMnAs film. The free energy for the magnetization oriented along a specific direction is given by the distance from the center of the coordinate system to the surface. The magnetic anisotropy in Fig.~\ref{fig:schema}(b) is dominated by a uniaxial anisotropy perpendicular to the film plane (the film plane is indicated by the orange plane). Consequently, the magnetization will be oriented within the film plane at low external fields in good approximation. In the film plane magnetic anisotropy mainly is determined by the cubic contribution with easy axes along $\left<100\right>$ at liquid He temperatures. 

To achieve a piezo-control of the magnetization orientation, the GaMnAs film is cemented onto the piezoelectric actuator (Sec.~\ref{sec:experimental}). Regarding the relative orientation of the GaMnAs film and the main elongation direction of the piezoelectric actuator there are two qualitatively different approaches: (\textit{i}) To achieve an irreversible switching of the magnetization from one minimum of cubic anisotropy into another, the GaMnAs film has to be affixed onto the piezoelectric actuator with an easy axis parallel to the main expansion direction of the actuator [Fig.~\ref{fig:schema}(a)]~\cite{Ove08}. In this way, starting from a GaMnAs film magnetized along the main expansion direction, the application of a positive voltage will lead to tensile strain and thus induce a uniaxial magnetoelastic contribution with a magnetic hard axis along the expansion direction [Fig.~\ref{fig:schema}(c)]. Thereby, the magnetization can be switched by $90^\circ$ into the direction of the new global minimum in free energy surface [Fig.~\ref{fig:schema}(d), Fig.~\ref{fig:schema2}(a)]. We note that two equivalent global minima are present at $\left[010\right]$ and $\left[0\overline{1}0\right]$. In order to lift this degeneracy, \textit{e.g.} a small external magnetic field should be applied in experiment. (\textit{ii}) In contrast, to be able to continuously and reversibly rotate magnetization within the film plane, we here will affix the GaMnAs film onto the piezoactor with the orientation of the [110] sample edge parallel to the main expansion direction (Fig.~\ref{fig:Experimental}). The difference between these two approaches can be visualized via cuts through the free energy surface within the film plane (Fig.~\ref{fig:schema2}). While in the alignment (\textit{i}) [Fig.~\ref{fig:schema2}(a)] the relative strength of the magnetic easy axes is changed, the alignment (\textit{ii}) [Fig.~\ref{fig:schema2}(b)] allows a rotation and thus a change of the relative orientation of the magnetic easy axes.

\section{\label{sec:experimental}EXPERIMENTAL}

\begin{figure}[tbp]
\includegraphics[width=0.3\textwidth]{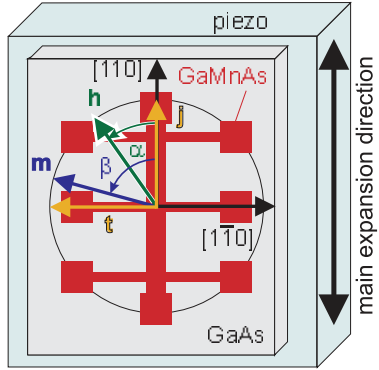} \vspace{-1.0cm}
\caption{(color online) Illustration of the relative alignment of the Hall bar, the GaMnAs thin film, and the main expansion direction of the piezoelectric actuator. $\textbf{h}=\textbf{H}/H$, $\textbf{m}=\textbf{M}/M$, $\textbf{j}$, and $\textbf{t}$ denote the unit vectors along the orientation of the magnetic field $H$, the magnetization $M$, the current density $j$, and the direction transverse to the current, respectively.}
\label{fig:Experimental}
\end{figure}

The 30~nm thick Ga$_{1-x}$Mn$_x$As film with a Mn concentration $x=0.045$ investigated here was grown on a (001)-oriented GaAs substrate by low-temperature molecular-beam epitaxy (LT-MBE) \cite{Lim06}. The sample exhibits a Curie temperature $T_{\mathrm{C}}\approx85$~K determined from superconducting quantum interference device (SQUID) magnetometry. Via optical lithography and wet chemical etching we patterned the film into Hall bars with the current direction \textbf{j} along the [110] crystal axis. After the etching process the GaAs substrate was mechanically polished to a thickness of about $100$~$\mathrm{\mu m}$. The sample then was cemented onto a lead zirconate titanate (PZT) piezoelectric actuator "PSt 150/2$\times$3/5a" (Piezomechanik M\"unchen) using the two-component epoxy "M-Bond 600" (Vishay Inc.) annealed for 2~h at 120~$^\circ$C in air. As shown in Fig.~\ref{fig:Experimental} the main piezoelectric actuator expansion direction is aligned parallel to the GaAs [110] direction.

We extracted the magnetic anisotropy from magnetotransport experiments carried out in a superconducting magnet cryostat. The sample was mounted on a rotatable sample stage that allows rotation around one axis, in such a way that the applied magnetic field always lay within the film plane, and that the angle $\alpha$ enclosed by the magnetic field orientation \textbf{h} and the current density $\mathbf{j}$ could be adjusted within $-140^\circ<\alpha<140^\circ$, as illustrated in Fig.~\ref{fig:Experimental}. Using a dc current density of $j=4.4\times10^3\mathrm{Acm^{-2}}$ the resistivities $\rho_{\mathrm{long}}$ along $\mathbf{j}$ and $\rho_{\mathrm{trans}}$ perpendicular to $\mathbf{j}$ were recorded via four-point measurements. We studied the magnetic properties of the sample using conventional magnetoresistance traces $\left\{\rho_{\rm long}(\mu_0H),\rho_{\rm trans}(\mu_0H)\right\}$ at a fixed magnetic field orientation $\alpha$, as well as  via angle-dependent magnetoresistance (ADMR) measurements $\left\{\rho_{\rm long}(\alpha),\rho_{\rm trans}(\alpha)\right\}$ at a fixed external magnetic field strength.

\section{\label{sec:strain}MEASUREMENT OF PIEZO-INDUCED STRAIN}

\begin{figure}[tbp]
\includegraphics[width=0.7\textwidth]{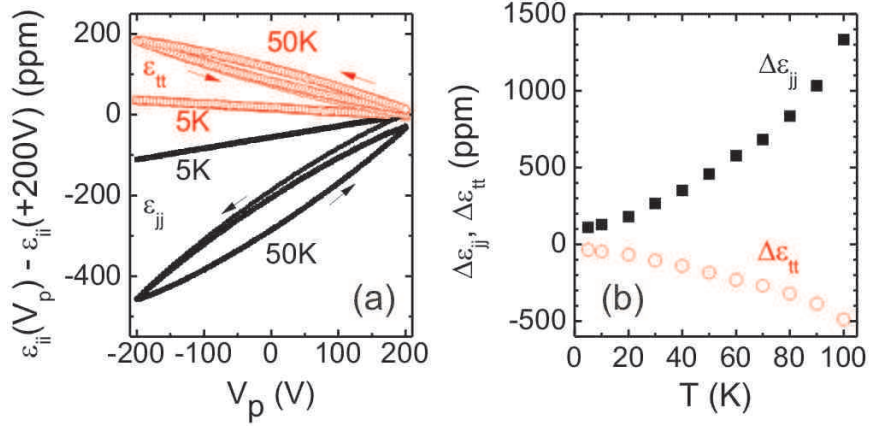} \vspace{-0.5cm}
\caption{(color online) (a) Longitudinal ($\epsilon_{jj}$, black full symbols) and transverse ($\epsilon_{tt}$, red open symbols) lattice distortions for piezo-voltage sweeps at 5~K (squares) and 50~K (circles) determined via strain gauges. The data show three full voltage cycles. To allow for comparison, we plot $\epsilon_{ii}\left(V_{\rm p}\right)-\epsilon_{ii}\left(+200~{\rm V}\right)$, $i\in\left\{j,t\right\}$.  (b) Temperature dependence of the piezo-induced lattice distortions $\Delta \epsilon_{ii}=\epsilon_{ii}(V_{\rm p}=+200~{\rm V})-\epsilon_{ii}(V_{\rm p}=-200~{\rm V})$, $i\in\left\{j,t\right\}$, determined from the third voltage cycle.}
\label{fig:epsilon}
\end{figure}

Application of a positive piezo-voltage results in a lattice elongation $\epsilon_{jj}$ along \textbf{j} and a lattice contraction $\epsilon_{tt}$ along \textbf{t} perpendicular to \textbf{j} (cf. Fig.~\ref{fig:Experimental}). To determine the temperature dependence of the piezo-induced lattice distortions $\epsilon_{ii}(V_{\rm p})$, $i\in\left\{j,t\right\}$ we used strain gauges. Figure~\ref{fig:epsilon}(a) depicts the measured longitudinal ($\epsilon_{jj}$) and transverse ($\epsilon_{tt}$) lattice distortions for piezo-voltage sweeps at 5~K and 50~K. For increasing temperature we observe an increase in both the total lattice distortions $\Delta \epsilon_{ii}=\epsilon_{ii}(V_{\rm p}=+200~{\rm V})-\epsilon_{ii}(V_{\rm p}=-200~{\rm V})$, $i\in\left\{j,t\right\}$ [Fig.~\ref{fig:epsilon}(b)] and the hysteresis exhibited by the piezoelectric actuator [Fig.~\ref{fig:epsilon}(a)].  This piezo-voltage induced strain causes an additional magnetoelastic contribution $F_{\rm magel}(V_{\rm p})/M=B_{110}(V_{\rm p})\cos^2\beta$ to the magnetic anisotropy as derived in the Appendix. Positive (negative) piezo-voltage thus results in an additional uniaxial magnetic anisotropy with a magnetic hard (easy) axis along \textbf{j}-direction.

\section{\label{sec:limmer}DETERMINATION OF MAGNETIC ANISOTROPY FROM MAGNETOTRANSPORT}

\begin{figure}[tbp]
\includegraphics[width=0.7\textwidth]{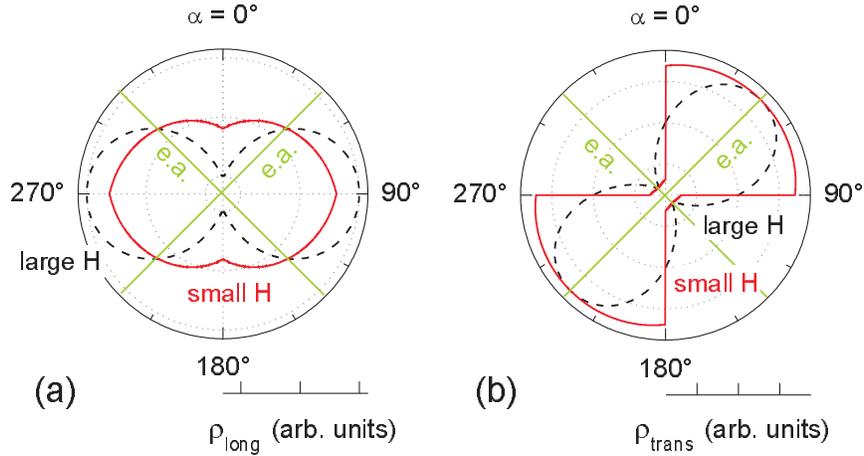} \vspace{-1.0cm}
\caption{(color online) Black dashed curves: Illustration of (a) longitudinal resistivity $\rho_{\rm long}(\alpha)$ and (b) transverse resistivity $\rho_{\rm trans}(\alpha)$ for a large external magnetic field $H$ rotated within the film plane. Since in this case the magnetization is aligned in field direction, \textit{i.e.} $\beta=\alpha$, the resistivities can be calculated from Eq. (\ref{long}) and (\ref{trans}). The parameters have been chosen with $\rho_0$ and $\rho_7>0$ and with $\rho_1$ and $\rho_3<0$. Red full curves: At small magnetic fields $H$, the magnetization tends to remain oriented close to a magnetic easy axis (e.a., green straight lines) and $\rho_{\rm long}$ and $\rho_{\rm trans}$ therefore tend to remain constant over a broad range of external magnetic field orientations nearby a magnetic easy axis.}
\label{fig:Figure1}
\end{figure}

We now describe the determination of the magnetic anisotropy from magnetotransport measurements in which the external magnetic field $H$ is rotated at different field strengths within the film plane as introduced above. Due to the dominating uniaxial anisotropy in growth direction caused by epitaxial strain, the magnetization in our film in good approximation is aligned within the film plane. Therefore, the longitudinal and the transverse resistivities
\begin{eqnarray}
\rho_{\rm long}&=&\rho_0+\rho_1(\textbf{j}\cdot\textbf{m})^2+\rho_3(\textbf{j}\cdot\textbf{m})^4=\rho_0+\rho_1\cos^2\beta + \rho_3 \cos^4\beta \label{long}\\
\rho_{\rm trans}&=&\rho_7(\textbf{t}\cdot\textbf{m})(\textbf{j}\cdot\textbf{m})=\frac{1}{2}\rho_7\sin(2\beta) \label{trans}
\end{eqnarray}
are determined by the angle $\beta$ between the in-plane orientation of the magnetization \textbf{m} and the current direction \textbf{j}.  The orientation \textbf{t} denotes the in-plane direction perpendicular to \textbf{j} (Fig.~\ref{fig:Experimental}). These expressions can be derived from a series expansion of the resistivity tensor in powers of the magnetization components $m_i$ \cite{Lim06}. Following Limmer \textit{et al.} \cite{Lim06} the resistivity parameters $\rho_0$, $\rho_1$, $\rho_3$, and $\rho_7$ can be deduced from magnetotransport measurements in which a constant large external magnetic field $H$ is rotated in the film plane. In good approximation, the magnetization in this case can be assumed to be oriented along the magnetic field direction, \textit{i.e.} $\beta=\alpha$ in Eq. (\ref{long}) and (\ref{trans}), where $\alpha$ denotes the angle between the magnetic field orientation \textbf{h} and the current direction \textbf{j}. Equations (\ref{long}) and (\ref{trans}) then yield the black dashed curves for $\left\{\rho_{\rm long}(\alpha),\rho_{\rm trans}(\alpha)\right\}$ in Fig.~\ref{fig:Figure1}. Note that angles are defined in the mathematically positive sense (counterclockwise) in Fig~\ref{fig:Experimental} and are plotted clockwise in the following. Thus $+90^\circ$ corresponds to the $\left[\overline{1}10\right]$ direction and not $\left[1\overline{1}0\right]$ as given in \cite{Goe08}. In contrast, for measurements at smaller external magnetic fields the orientation of magnetization $-$ and also $\rho_{\rm long}$ and $\rho_{\rm trans}$ $-$ to an increasing extent will be influenced by the magnetic anisotropy. Therefore, at small magnetic fields the magnetization will tend to remain oriented close to a magnetic easy axis (e.a., depicted as green straight lines in Fig.~\ref{fig:Figure1}) $-$ and $\rho_{\rm long}$ and $\rho_{\rm trans}$ will tend to remain constant $-$ over a broad range of external magnetic field orientations near to a magnetic easy axis. Accordingly, abrupt changes in $\left\{\rho_{\rm long}(\alpha),\rho_{\rm trans}(\alpha)\right\}$ indicate a nearby magnetic hard axis, as shown by the red full curves in Fig.~\ref{fig:Figure1}. Note that a similar approach to determine the in-plane magnetic anisotropy in GaMnAs has recently been followed by Yamada \textit{et al.}~\cite{Yamada06}.

\begin{figure}[tbp]
\includegraphics[width=0.7\textwidth]{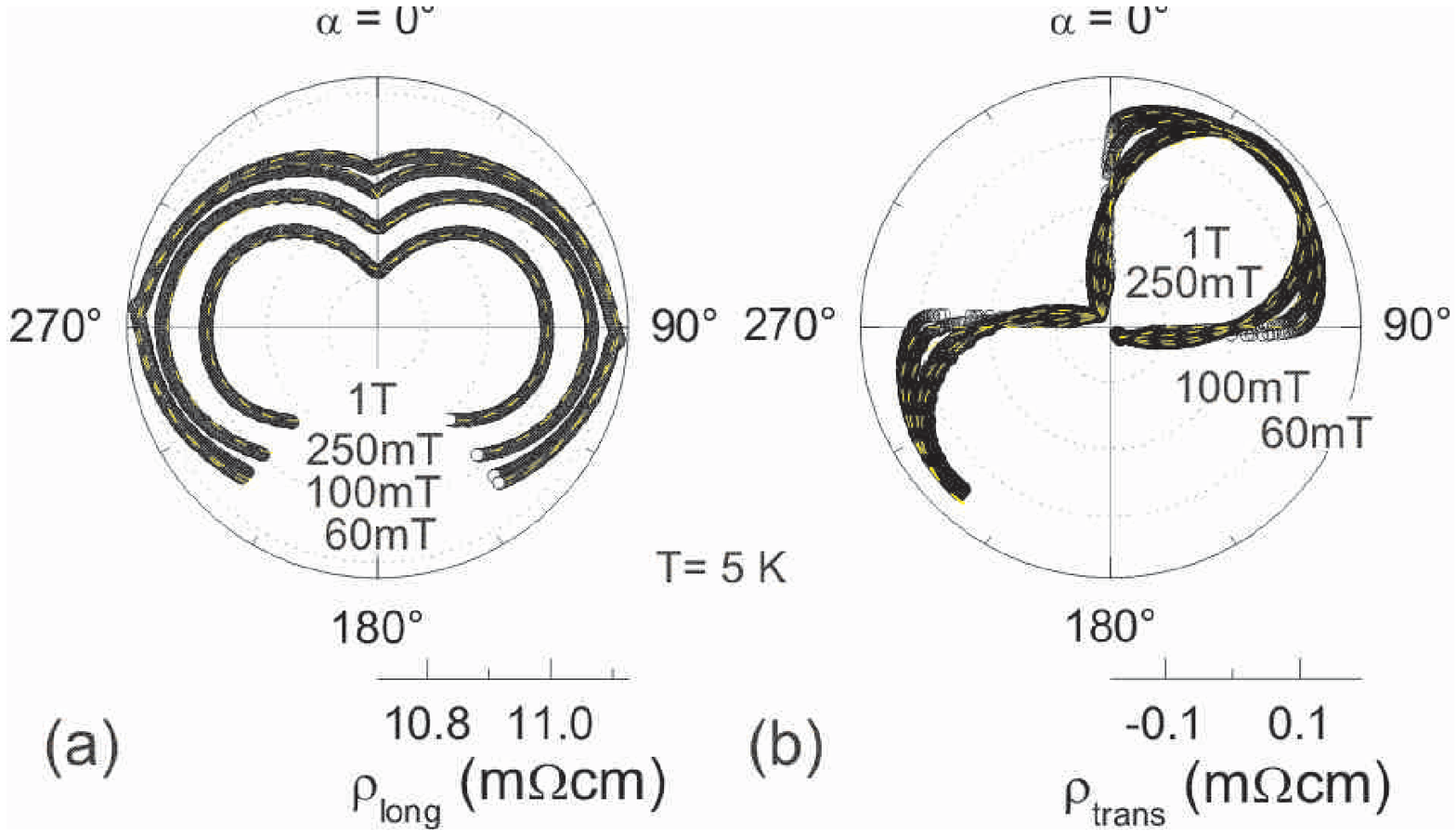} 
\caption{(color online) (a) Longitudinal resistivity $\rho_{\rm long}(\alpha)$ and (b) transverse resistivity $\rho_{\rm trans}(\alpha)$ for a constant external magnetic field $\mu_0H=60$~mT, 100~mT, 250~mT, and 1~T rotated within the film plane for $V_{\rm p}=0$~V. The black circles represent the experimental data, the yellow dashed curves the result of the corresponding simulation using the parameter values given in Fig.~\ref{fig:Figure5}.}
\label{fig:Figure2}
\end{figure}

For the angle-dependent magnetoresistance (ADMR) measurements we first aligned the magnetization into a well-defined initial state in an external magnetic field of $\mu_0H=7$~T along $\alpha=-140^\circ$. Then, we lowered the field to the measurement field and started the angular scan $\left\{\rho_{\rm long}(\alpha),\rho_{\rm trans}(\alpha)\right\}$. Figure~\ref{fig:Figure2} shows $\left\{\rho_{\rm long}(\alpha),\rho_{\rm trans}(\alpha)\right\}$ measured at $T=5$~K and $V_{\rm p}=0$~V for rotations within external magnetic fields $\mu_0H$ of 1~T, 250~mT, 100~mT, and 60~mT. Following the discussion above, the angular dependence at small $\mu_0H$ already indicates that the in-plane anisotropy is dominated by a cubic contribution with magnetic easy axes along $\left\langle 100\right\rangle$. This is corroborated by the simulation, where the simulated curves (yellow dashed) are the result of a procedure in which the resistivity parameters and the anisotropy parameters from Eq.~(\ref{eq:F}) are fitted iteratively~\cite{Lim06}. The values for $\rho_0$, $\rho_1$, $\rho_3$, and $\rho_7$ and the anisotropy fields thus obtained are summarized in Fig.~\ref{fig:Figure5} below.  We note that in this fit all resistivity parameters except for $\rho_0$ are kept constant for the different values of the applied magnetic field. The change in $\rho_0$ with magnetic field strength accounts for the influence of negative magnetoresistance typically observed for GaMnAs~\cite{Goe05}.

\section{\label{sec:switching5K}ADMR AND PIEZO-STRAIN AT 5~K}

At 5~K the influence of piezo-induced strain on the ADMR scans $\left\{\rho_{\rm long}(\alpha),\rho_{\rm trans}(\alpha)\right\}$ discussed in the preceding section is only marginal. This is due to the small piezo-induced lattice distortion at 5~K (Fig.~\ref{fig:epsilon}) and the large cubic anisotropy at this temperature. However, the influence of the magnetoelastic contribution can be resolved in conventional $\left\{\rho_{\rm long}(\mu_0H),\rho_{\rm trans}(\mu_0H)\right\}$ magnetotransport measurements.

\begin{figure}[tbp]
\includegraphics[width=0.7\textwidth]{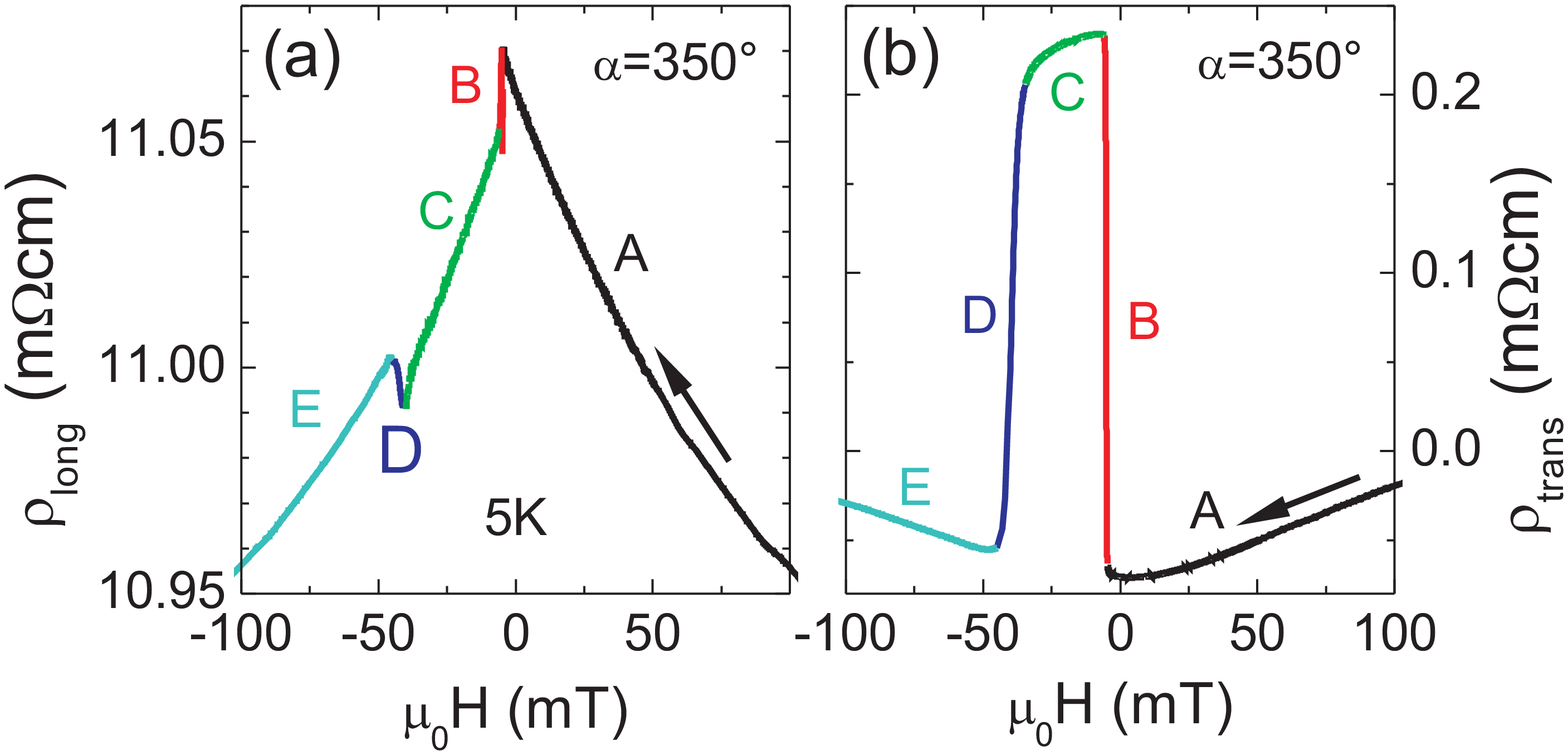} \vspace{-2.5cm}
\caption{(color online) (a) $\rho_{\rm long}(\mu_0H)$ and (b) $\rho_{\rm trans}(\mu_0H)$ curves measured at 5~K for the external magnetic field $H$ aligned at $\alpha=350^\circ$. An explanation of the evolution is given in the text.}
\label{fig:Figure7}
\end{figure}

Figure~\ref{fig:Figure7} shows the $\left\{\rho_{\rm long}(\mu_0H),\rho_{\rm trans}(\mu_0H)\right\}$ downsweep curves measured at 5~K for the external magnetic field $H$ aligned at $\alpha=350^\circ$. The evolution of the downsweep curve for $V_{\rm p}=0$~V from positive to negative magnetic fields can be understood as follows (compare the corresponding evolution of $\left\{\rho_{\rm long}(\alpha),\rho_{\rm trans}(\alpha)\right\}$ in Fig.~\ref{fig:Figure2} at $\mu_0H=1$~T where $\beta=\alpha$ is fulfilled): (A) $\mu_0H=+100$~mT$\rightarrow -5$~mT: rotation of magnetization from $\beta=350^\circ \rightarrow 315^\circ$, the orientation of the closest magnetic easy axis. (B) $\mu_0H=-5$~mT: first switching process by $\Delta \beta\approx90^\circ$ into a direction close to the magnetic easy axis at $225^\circ$. We refer to the corresponding field as a switching field in the following. (C) $\mu_0H=-5$~mT$\rightarrow-40$~mT: slight rotation of the magnetization (corresponding to a decrease of $\beta$) within the minimum of free energy close to $225^\circ$. (D) $\mu_0H=-40$~mT: second switching process by $\Delta \beta\approx90^\circ$ into a direction close to the magnetic easy axis at $135^\circ$. (E) $\mu_0H=-40$~mT$\rightarrow-100$~mT: rotation of magnetization from $\beta\approx135^\circ \rightarrow 170^\circ$, the orientation of the external magnetic field. Note that the change in $\left\{\rho_{\rm long}(\mu_0H),\rho_{\rm trans}(\mu_0H)\right\}$ during each step (A to E) of the downsweep described above is in agreement with the values expected for the corresponding orientation of magnetization from the angle-dependent resistivity loops  $\left\{\rho_{\rm long}(\alpha),\rho_{\rm trans}(\alpha)\right\}$ at high external magnetic field in Fig.~\ref{fig:Figure2}, where $\beta=\alpha$ is fulfilled. For example, this comparison explains the fact that the jumps observed in $\rho_{\rm trans}$ are much larger than the ones observed in $\rho_{\rm long}$, since for the first switching process (B) $\rho_{\rm trans}$ jumps from its minimum value at $315^\circ$ to its maximum value at $225^\circ$, while $\rho_{\rm long}$ exhibits the same values at $315^\circ$ and  $225^\circ$. Due to the larger resistivity difference in $\rho_{\rm trans}$ in the following we will restrict our discussion to this quantity. Note that we nevertheless observe a consistent behavior for $\rho_{\rm long}$.

\begin{figure}[tbp]
\includegraphics[width=0.65\textwidth]{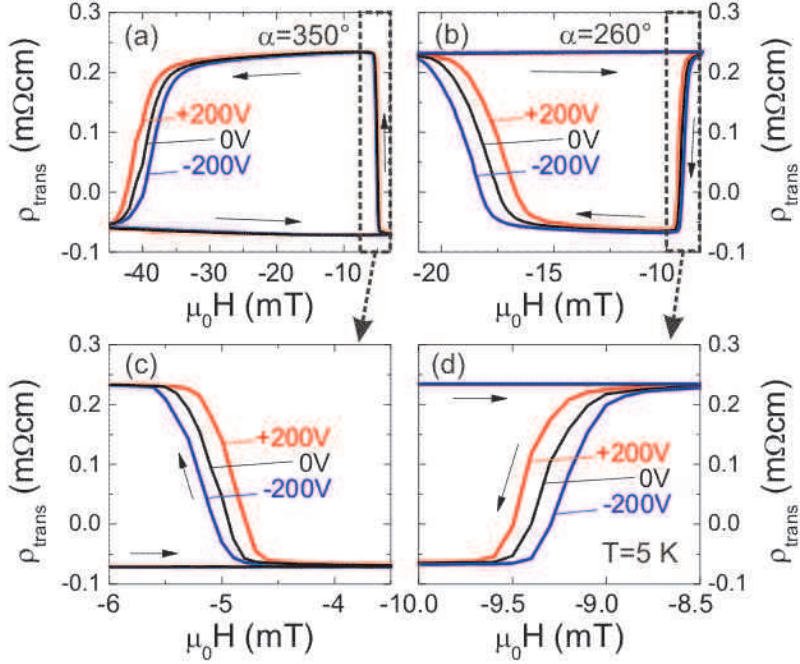} 
\caption{(color online) (a),(c) Enlarged sections of the $\rho_{\rm trans}(\mu_0H)$ curves measured at 5~K for the external magnetic field $H$ aligned at $\alpha=350^\circ$ showing the piezo-voltage dependence of the first (c) and second (a) switching field. (b),(d) For $\alpha=260^\circ$ the order of switching fields depending on the piezo voltage is inverted. The arrows indicate the direction of the field sweeps.}
\label{fig:Figure10}
\end{figure}
Figure~\ref{fig:Figure10}(c),(a) shows the dependence of the first (B) and the second (D) switching field on $V_{\rm p}$ for the same magnetic field orientation $\alpha=350^\circ$. For positive (negative) $V_{\rm p}$ the first switching field shifts to lower (higher) and the second switching field to higher (lower) absolute values of the magnetic field. This observation can be explained via a slight tilt of the relative orientation of the magnetic easy axes. A positive $V_{\rm p}$ induces a uniaxial magnetic anisotropy with a magnetic hard axis along \textbf{j} leading to a switching angle $<90^\circ$ for the first and $>90^\circ$ for the second switching process. The smaller the switching angle, the smaller is the energy cost for the domain wall formation around a new magnetic domain nucleus. The latter has to be accounted for by the energy gain of the magnetization switching into the energetically more favorable minimum in free energy. Therefore, a smaller (larger) angle between two magnetic easy axes leads to a shift of the corresponding switching field to smaller (larger) absolute values of the magnetic field, in agreement with experimental observation.
For the external magnetic field aligned along $\alpha=350^\circ-90^\circ=260^\circ$ the order of the switching fields is inverted compared to $\alpha=350^\circ$ [Figure~\ref{fig:Figure10}(b),(d)]. This is due to the fact that the order of the switching angles is inverted, too, corroborating our analysis.

\section{\label{sec:50K}ADMR AND PIEZO-STRAIN AT 50~K}

\begin{figure}[tbp]
\includegraphics[width=0.55\textwidth]{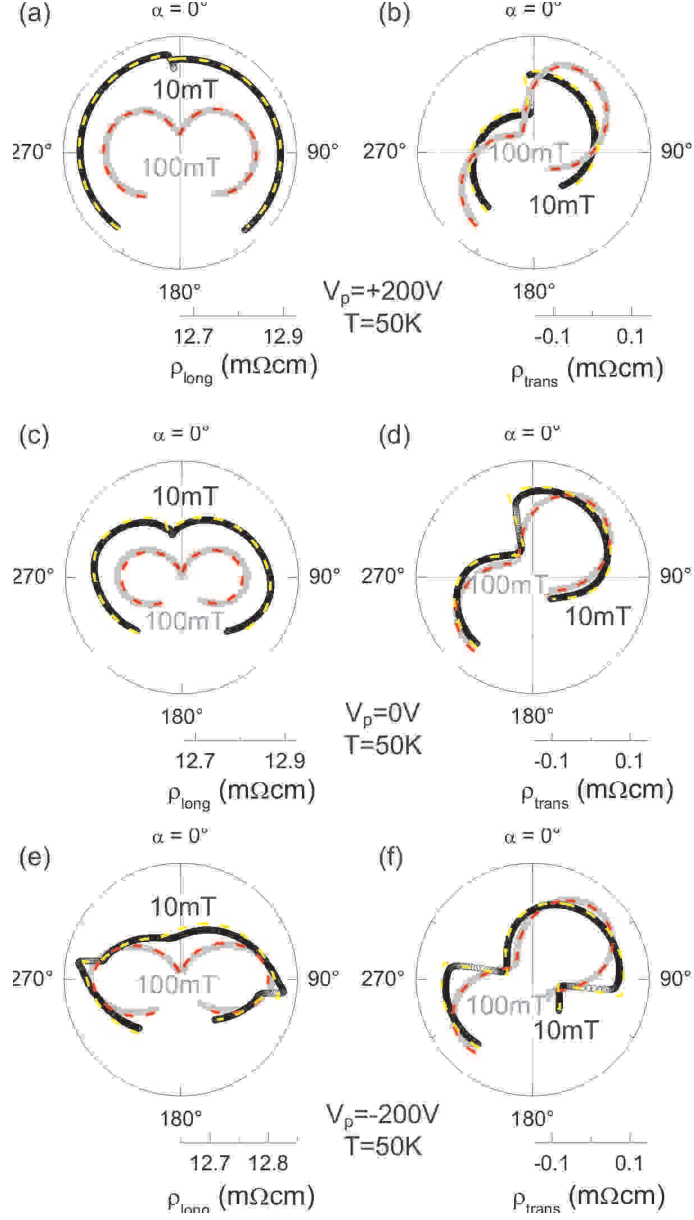} 
\caption{(color online) Longitudinal resistivity $\rho_{\rm long}(\alpha)$ and transverse resistivity $\rho_{\rm trans}(\alpha)$ at 50~K for a constant external magnetic field rotated within the film plane at different voltages $V_{\rm p}$ applied to the piezoelectric actuator: (a),(b) $V_{\rm p}=+200V$, (c),(d)  $V_{\rm p}=0V$, and (e),(f)  $V_{\rm p}=-200V$. The black and grey symbols correspond to the ADMR curves obtained at $\mu_0H=10$~mT and $\mu_0H=100$~mT, respectively. The yellow ($\mu_0H=10$~mT ) and red ($\mu_0H=100$~mT) dashed curves are obtained from the iterative fitting procedure of the ADMR curves at different external magnetic field strengths using the parameter values given in Fig.~\ref{fig:Figure5}.}
\label{fig:Figure3}
\end{figure}

Due to the strong temperature dependence of magnetic anisotropy in GaMnAs (Sec.~\ref{sec:introduction} and \ref{sec:lambda}) the magnetic anisotropy landscape can be adjusted to a regime in which piezo-voltage induced strain qualitatively alteres magnetic anisotropy~\cite{Goe08}. This becomes obvious from the $\left\{\rho_{\rm long}(\alpha),\rho_{\rm trans}(\alpha)\right\}$ ADMR measurements at 50~K (Fig.~\ref{fig:Figure3}). For $\mu_0H=100$~mT, the magnetization orientation is still predominantly determined by the orientation of the external magnetic field $\alpha$. Therefore, at $\mu_0H=100$~mT the $\left\{\rho_{\rm long}(\alpha),\rho_{\rm trans}(\alpha)\right\}$ curves at different piezo voltages only exhibit minor changes. However, for $\mu_0H=10$~mT, the magnetization orientation and thus also the magnetic anisotropy is strongly influenced by the piezo-induced strain, which can be deduced from the qualitatively different behavior of $\left\{\rho_{\rm long}(\alpha),\rho_{\rm trans}(\alpha)\right\}$ at different piezo voltages. As shown in Fig.~\ref{fig:Figure3}(a),(b) for an external magnetic field of 10~mT at $V_{\rm p}=+200$~V, $\rho_{\rm long}$ and $\rho_{\rm trans}$ exhibit abrupt changes for an angle of the magnetic field $\alpha\approx0^\circ$ between \textbf{h} and \textbf{j}, indicating a nearby magnetic hard axis. The smooth changes in resistivity around $\alpha\approx90^\circ$ indicate a nearby magnetic easy axis. In contrast, the curvature for $V_{\rm p}=-200$~V [Fig.~\ref{fig:Figure3}(e),(f)] evolves approximately vice versa, indicating a rotation of the magnetic easy axis by almost 90$^\circ$. In a more thorough analysis the anisotropy parameters are determined independently for each piezo voltage via the iterative fitting procedure of the ADMR curves at different external magnetic field strengths as already described above for the measurements at 5~K (Sec.~\ref{sec:limmer}). In this way we are able to quantitatively determine the piezo-voltage induced change in the free energy surface at $T=50$~K. This analysis yields a rotation of the magnetic easy axis by about 70$^\circ$.

\section{\label{sec:rotation}REVERSIBLE PIEZO-VOLTAGE CONTROL OF MAGNETIZATION ORIENTATION}

\begin{figure}[tbp]
\includegraphics[width=0.5\textwidth]{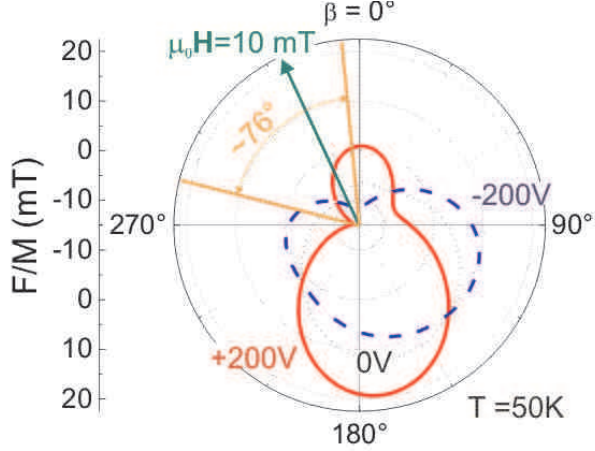} 
\caption{(color online) Free energy normalized to the saturation magnetization $F/M(\beta)$ as a function of magnetization orientation $\beta$ for an external magnetic field $\mu_0H=10$~mT applied along $\alpha=-25^\circ$ taking into account the magnetic anisotropy parameters derived from the ADMR at different magnetic field strengths at 50~K depicted in Fig.~\ref{fig:Figure3}. The magnetization orientation is expected to change by $\Delta \beta = 76^\circ$ when the piezo voltage $V_{\rm p}$ is varied between $+200$~V and $-200$~V.}
\label{fig:Figure4_2}
\end{figure}

As discussed in the preceding section we are able to rotate the orientation of the magnetic easy axis by about 70$^\circ$ via application of different piezo-voltages at 50~K. However, since at zero external magnetic field the magnetization can decay into magnetic domains aligned either parallel or antiparallel to the magnetic easy axis, a piezo-voltage control of magnetization orientation is not granted. To prevent this issue of domain formation, we apply a small external magnetic field fixed at $\alpha=-25^\circ$, which lifts the degeneracy between antiparallel and parallel alignments of \textbf{m} relative to the magnetic easy axis. In this way only one global free energy minimum prevails, which is oriented along $\beta=-80^\circ$ for $V_{\rm p}=+200$~V and along $\beta=-5^\circ$ for $V_{\rm p}=-200$~V for $\mu_0H=10$~mT (Fig.~\ref{fig:Figure4_2}).

\begin{figure}[tbp]
\includegraphics[width=0.55\textwidth]{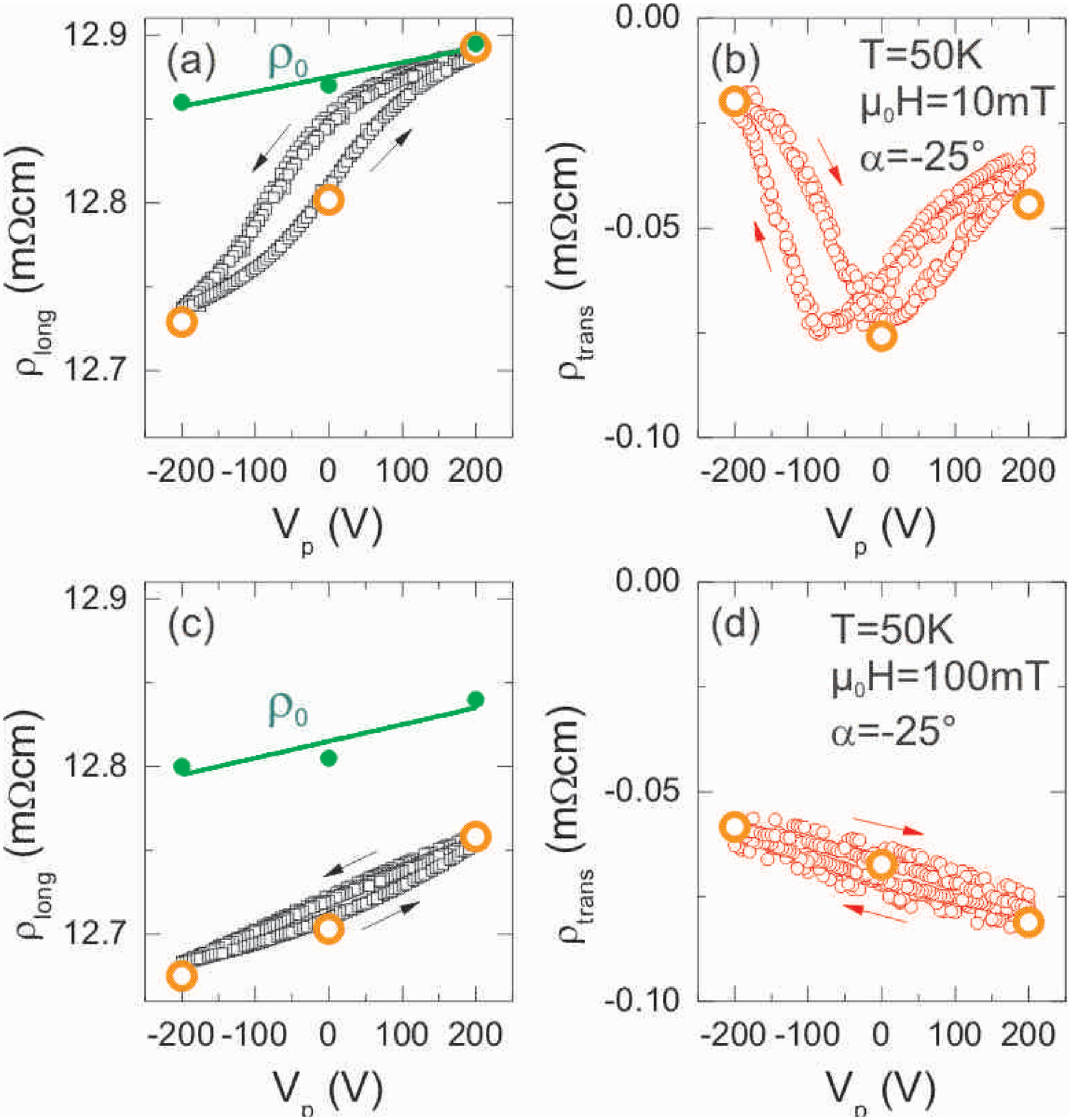} 
\caption{(color online) Evolution of (a) $\rho_{\rm long}(V_{\rm p})$ and (b) $\rho_{\rm trans}(V_{\rm p})$ as a function of the variation of the piezo voltage $V_{\rm p}$ between $+200$~V and $-200$~V for $\mu_0H=10$~mT applied along $\alpha=-25^\circ$ at 50~K. The solid green line in (a) displays the elongation dependence of the resistivity parameter $\rho_0$ in $\rho_{\rm long}$ [compare Eq. (\ref{long})] as derived from the corresponding ADMR curves, which is subtracted from $\rho_{\rm long}$ for the magnetization orientation determination. (c),(d) Corresponding measurements for $\mu_0H=100$~mT. In all panels the orange open circles at $V_{\rm p}=-200$~V, 0~V, and +200~V display the resistivity values obtained from the ADMR measurements at $\alpha=-25^\circ$ (cf. Fig.~\ref{fig:Figure3}), demonstrating the good reproducibility.}
\label{fig:Figure8}
\end{figure}

We now take advantage of the direct correspondence between $\left\{\rho_{\rm long},\rho_{\rm trans}\right\}$ and the magnetization orientation to \textit{in situ} monitor the evolution of magnetization orientation as a function of $V_{\rm p}$.  A measurement of $\rho_{\rm long}(V_{\rm p})$ and $\rho_{\rm trans}(V_{\rm p})$ allows to recalculate the corresponding magnetization orientation $\beta(V_{\rm p})$ via inversion of Eq.~(\ref{long}) and (\ref{trans}). For details regarding the inversion we refer to Ref.~\cite{Goe08}. Since the data obtained for $\mu_0H=10$~mT applied along $\alpha=-25^\circ$ have already been discussed in detail in \cite{Goe08}, in the following we will focus on the changes observed for different external magnetic field strengths. Figure~\ref{fig:Figure8}(a),(b) [(c),(d)] shows $\rho_{\rm long}(V_{\rm p})$ and $\rho_{\rm trans}(V_{\rm p})$ at $T=50$~K for a constant field $\mu_0H=10$~mT [$\mu_0H=100$~mT] applied along $\alpha=-25^\circ$. The data consist of three full voltage cycles ($V_{\rm p}=+200$~V$\rightarrow -200$~V$\rightarrow+200$~V) in 5~V steps, evidencing the good reproducibility of the measurement. The hysteresis is due to the hysteretic expansion/contraction of the piezoelectric actuator with $V_{\rm p}$ [Fig.~\ref{fig:epsilon}(a)].

\begin{figure}[tbp]
\includegraphics[width=0.35\textwidth]{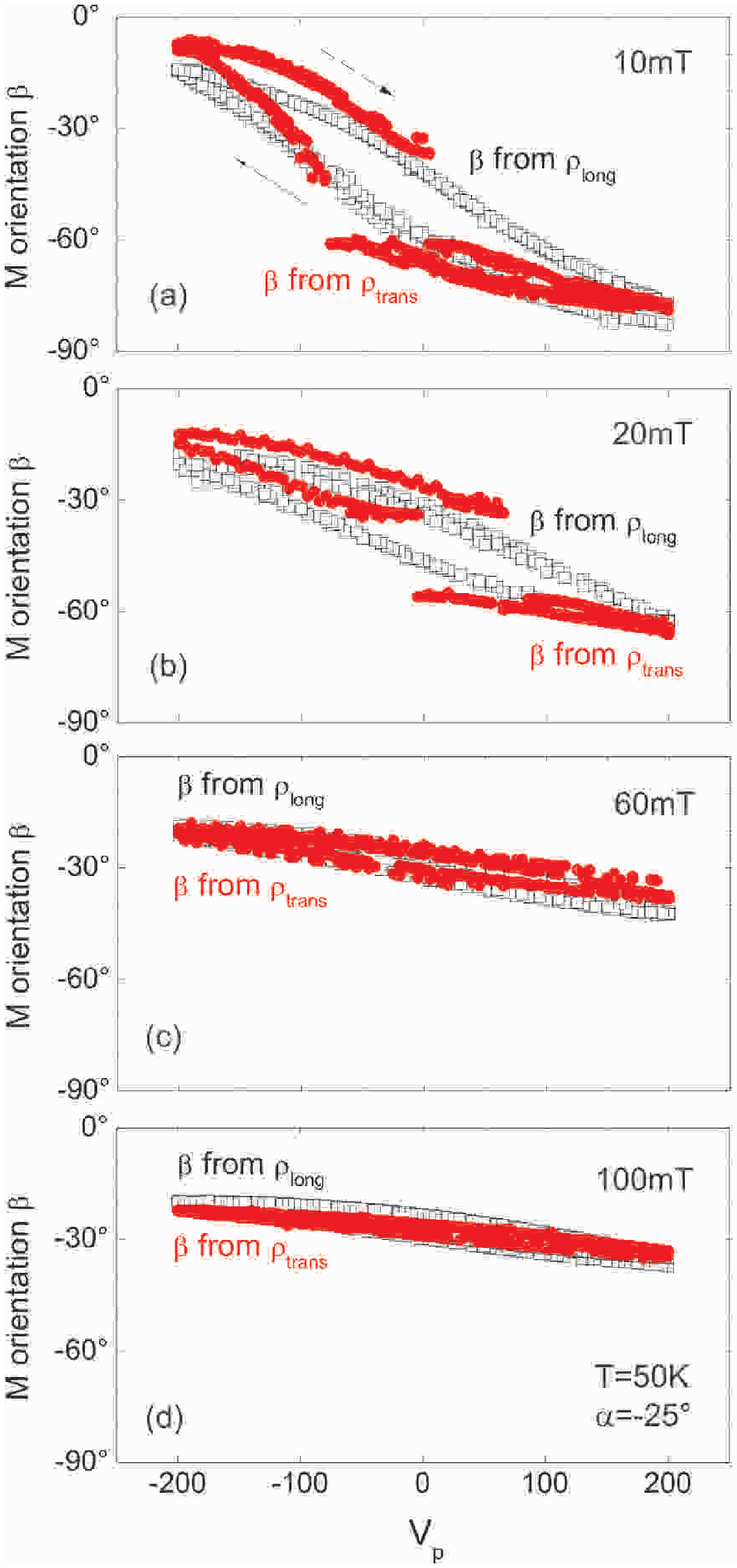} 
\caption{(color online) The magnetization orientation $\beta$ as a function of piezo voltage $V_{\rm p}$ at 50~K for different external magnetic field strengths $\mu_0H$ applied along $\alpha=-25^\circ$ calculated from $\rho_{\rm long}$ via Eq. (\ref{long}) (black open squares) and $\rho_{\rm trans}$ via Eq. (\ref{trans}) (red full circles). 
The good agreement between the $\beta$ values calculated from $\rho_{\rm long}$  and $\rho_{\rm trans}$ corroborates our analysis and demonstrates that for $\mu_0H=10$~mT the magnetization orientation can be continuously and reversibly rotated by about $\Delta \beta \approx 70^\circ$ solely by varying the piezo voltage between $+200$~V and $-200$~V, as expected from the free energy contours (Fig.~\ref{fig:Figure4_2}). For increasing $\mu_0H$ the maximum angle of rotation is decreased to $\Delta \beta \approx 15^\circ$ for $\mu_0H=100$~mT.}
\label{fig:resultUmrechnungPhiM}
\end{figure}

Figure \ref{fig:resultUmrechnungPhiM} shows the magnetization orientation $\beta(V_{\rm p})$ as a function of piezo voltage $V_{\rm p}$ at 50~K obtained from the $\rho_{\rm long}$ data (black open squares) and the $\rho_{\rm trans}$ data (red full circles) for different external magnetic fields strengths $\mu_0H$ applied along $\alpha=-25^\circ$. The abrupt jumps in $\rho_{\rm trans}$ at $V_{\rm p}\approx-75$~V and $V_{\rm p}\approx0$~V for $\mu_0H=10$~mT and 20~mT are an artefact of the arcsine transformation caused by the fact that the minimal value of $\rho_{\rm trans}(V_{\rm p}\approx 0~{\rm V})=-0.07$~m$\Omega$cm in Fig.~\ref{fig:Figure8}(b) is somewhat larger than the value $\frac{1}{2}\rho_7=-0.1$~m$\Omega$cm in Eq.~(\ref{trans}) determined from ADMR traces at 50~K (cf. Fig.~\ref{fig:Figure5}). The magnetization orientations $\beta(V_{\rm p})$ obtained from $\rho_{\rm long}$ and $\rho_{\rm trans}$ consequently are consistent and demonstrate that for $\mu_0H=10$~mT we indeed are able to continuously and reversibly adjust the magnetization orientation at will within about $70^\circ$ solely via application of a voltage to the piezoelectric actuator \cite{Goe08}. With increasing magnetic field strength, the Zeeman contribution in Eq.~(\ref{eq:F}) increasingly dominates over the magnetoelastic contribution to the free energy. Thus the angular range within which the magnetization orientation can be rotated as a function of $V_{\rm p}$ decreases from $\Delta \alpha \approx 70^\circ$ for $\mu_0H=10$~mT to $\Delta \alpha \approx 15^\circ$ for $\mu_0H=100$~mT.

\section{\label{sec:lambda}TEMPERATURE DEPENDENCE OF THE MAGNETIC ANISOTROPY AND THE MAGNETOSTRICTIVE CONSTANT}

\begin{figure}[tbp]
\includegraphics[width=0.55\textwidth]{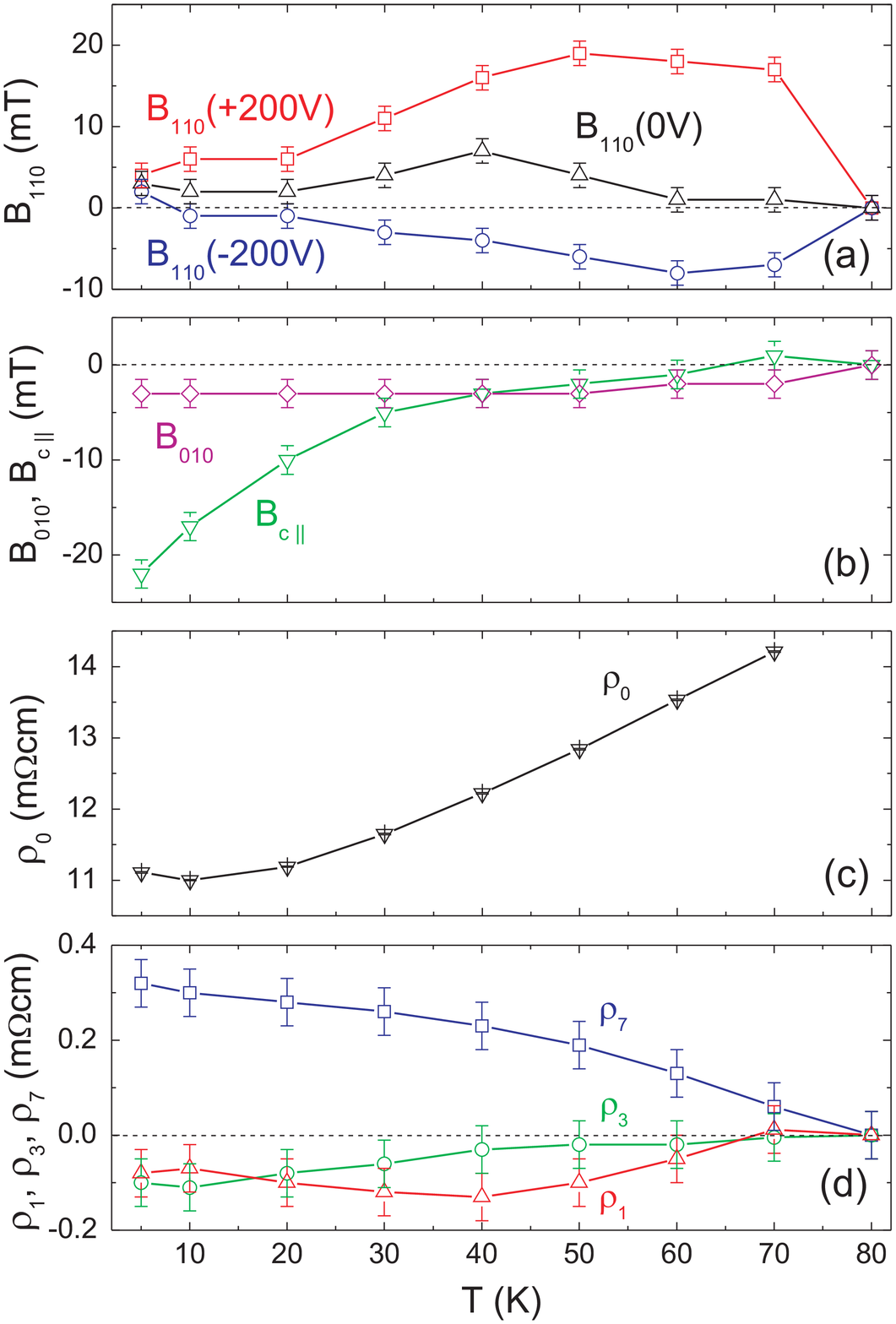} 
\caption{(color online) Temperature dependence of the anisotropy parameters $B_{110}(V_{\rm p})$, $B_{010}$, and $B_{\rm c||}$ (a), (b) and the resistivity parameters  $\rho_0$, $\rho_1$, $\rho_3$, and $\rho_7$ (c), (d) as derived from the iterative fitting procedure of the ADMR curves at different external magnetic field strengths. At temperatures $T\lesssim T_{\rm C}/2\approx40$~K the magnetic anisotropy is dominated by the cubic contribution, while at higher temperatures $T\gtrsim T_{\rm C}/2$ the piezo-voltage induced uniaxial magnetic anisotropy contribution becomes more important. Parameter $\rho_0$ in (c) was derived from measurements at $V_{\rm p}=+200$~V and $\mu_0H=100$~mT, the parameters $\rho_1$, $\rho_3$, and $\rho_7$ are found to be independent of $V_{\rm p}$ within error bars.}
\label{fig:Figure5}
\end{figure}

\begin{figure}[tbp]
\includegraphics[width=0.7\textwidth]{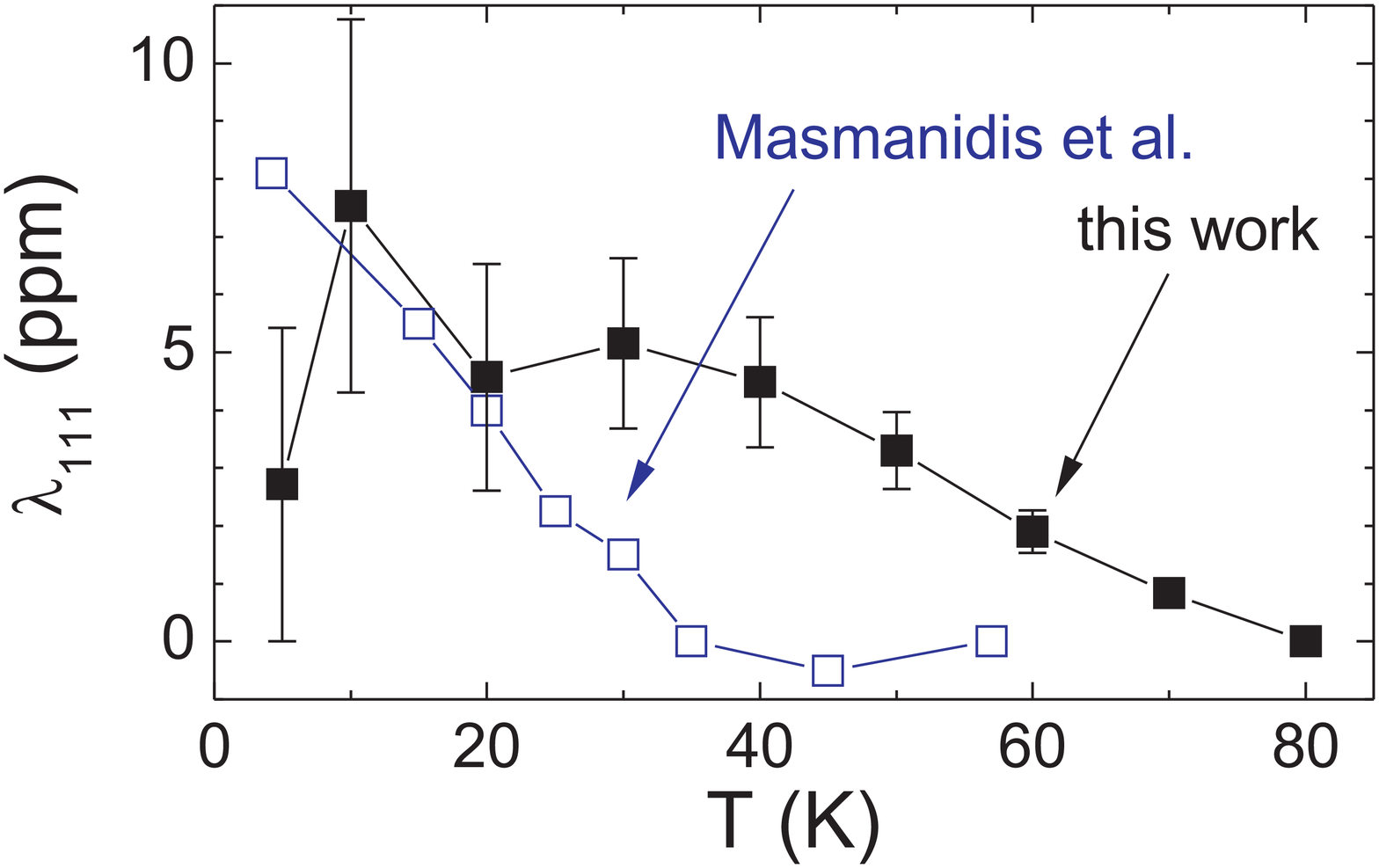} \vspace{-1.0cm}
\caption{(color online) Temperature dependence of the magnetostrictive constant (full black squares) compared to the values obtained by Masmanidis \textit{et al.} \cite{Mas05} via a nanoelectromechanical resonator (open blue squares).}
\label{fig:Figure6}
\end{figure}

From the full set of ADMR $\left\{\rho_{\rm long}(\alpha),\rho_{\rm trans}(\alpha)\right\}$ measurements and the corresponding analyses for $V_p\in\left\{+200~\rm V, 0~\rm V, -200~\rm V\right\}$ at different temperatures (Sec.~\ref{sec:50K}), we obtain the temperature dependence of the resistivity parameters $\rho_0$, $\rho_1$, $\rho_3$, and $\rho_7$ and the anisotropy parameters $B_{110}(V_{\rm p})$, $B_{010}$, and $B_{\rm c||}$ shown in Fig.~\ref{fig:Figure5}. Within error bars we did not observe a dependence of $\rho_1$, $\rho_3$, and $\rho_7$ on $V_{\rm p}$. The dependence of the resistivity parameters on the external magnetic field has recently been investigated by Wu \textit{et al.}~\cite{Wu08} in the range 0.5~T$\leq\mu_0H\leq9$~T. For the magnetic field range $\mu_0H\leq1$~T of our ADMR measurements we also did not observe a magnetic field dependence of $\rho_1$, $\rho_3$, and $\rho_7$, which within error bars is in agreement with Ref.~\cite{Wu08}. The piezo voltage dependence of $\rho_0$ at 50~K is exemplarily shown in Fig.~\ref{fig:Figure8}(a),(c).

As typically observed for GaMnAs, the magnetic anisotropy at low temperatures ($T\lesssim T_{\rm C}/2\approx40$~K) is dominated by the cubic magnetic anisotropy contribution, while at higher temperatures ($T\gtrsim T_{\rm C}/2$) the uniaxial magnetic anisotropy contributions gain importance~\cite{Ham03}. The strong increase of the magnetoelastic contribution $B_{110}(\pm200~\rm V)$ in the temperature range $25~\rm K\lesssim T\lesssim 50$~K is due to the increase of absolute piezo elongation with temperature (Fig.~\ref{fig:epsilon}).

In the following we use the magnetic anisotropy constants $B_{c ||}$, $B_{010}$ and $B_{110}$ as derived from the analysis of the magnetotransport measurements (Fig.~\ref{fig:Figure5}) and the piezo-induced lattice distortion (Fig.~\ref{fig:epsilon}) to derive the magnetostrictive constant $\lambda_{111}$. In contrast to $B_{110}$, this quantity is a property of the GaMnAs film and not dependent on the particular method used to apply stress. As detailed in the Appendix, the magnetostrictive constant $\lambda_{111}(T)$ can be calculated from
\begin{equation}
\lambda_{111}(T)=\frac{2 \Delta B_{110}(T) M(T)}{3 c_{44} \left[\Delta \epsilon_{jj}(T)-\Delta \epsilon_{tt}(T)\right]},
\label{equ:lambda}
\end{equation}
using the temperature dependence of the piezo-induced lattice distortions $\Delta \epsilon_{ii}=\epsilon_{ii}(V_{\rm p}=+200~{\rm V})-\epsilon_{ii}(V_{\rm p}=-200~{\rm V})$, $i\in\left\{j,t\right\}$ (Fig.~\ref{fig:epsilon}), of  $\Delta B_{110}(T)=B_{110}(T,V_{\rm p}=+200~{\rm V})-B_{110}(T,V_{\rm p}=-200~{\rm V})$ [Fig.~\ref{fig:Figure5}(a)], and of the magnetization $M(T)$ determined via superconducting quantum interference device (SQUID) magnetometry at a fixed magnetic field strength of $\mu_0H=100$~mT. For $c_{44}$, we use the elastic modulus of bulk GaAs. Since it only varies slightly with temperature, a fixed value of $c_{44}=59.9$~GPa~\cite{Bla82} is used in all calculations. The temperature dependence observed is summarized in Fig.~\ref{fig:Figure6} and compared to the results of Masmanidis \textit{et al.}~\cite{Mas05}. Around 10~K to 20~K there is good agreement with the values obtained via nanoelectromechanical measurements. Due to the small magnetoelastic contribution to the magnetic anisotropy at 5~K, we do not attribute a high significance to our data point at this low temperature. The deviation at higher temperatures between our data and Ref.~\cite{Mas05} can be accounted for by the differences in $T_{\rm C}$ between the 80~nm thick Ga$_{0.948}$Mn$_{0.052}$As sample in Ref.~\cite{Mas05} with $T_{\rm C}=57$~K and our 30~nm thick Ga$_{0.955}$Mn$_{0.045}$As sample with a much higher $T_{\rm C}=85$~K.

\section{\label{sec:NVswitching}PIEZO-VOLTAGE INDUCED NONVOLATILE MAGNETIZATION SWITCHING}

\begin{figure}[tbp]
\includegraphics[width=0.65\textwidth]{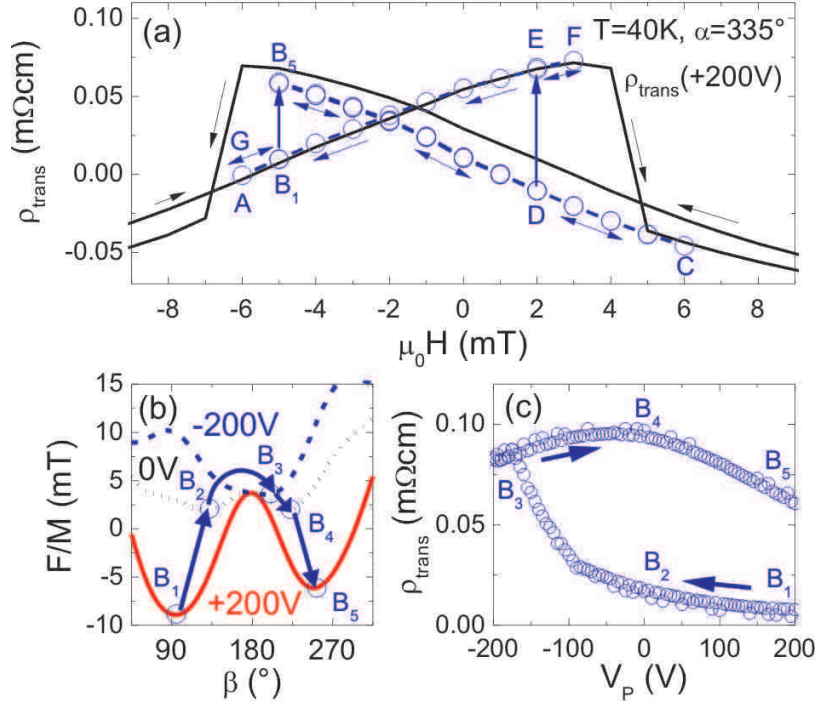} 
\caption{(color online) (a) Transverse resistivity loop $\rho_{\rm trans}(\mu_0H)$ at 40~K with $V_{\rm p}=+200$~V and the external magnetic field oriented along $\alpha=335^\circ$. The open blue circles describe the evolution of $\rho_{\rm trans}(\mu_0H)$ during the measurements sequence $A$ to $G$ described in the text. (b) Piezo-voltage induced change in free energy normalized on saturation magnetization $F/M(\beta)$ as a function of magnetization orientation $\beta$ for $\mu_0H=-5$~mT. (c) Evolution of $\rho_{\rm trans}$ during the piezo-voltage sweep from $V_{\rm p}=+200$~V ($B_1$) $\rightarrow 0$~V ($B_2$) $\rightarrow -200$~V ($B_3$)$\rightarrow0$~V ($B_4$)$\rightarrow +200$~V ($B_5$).}
\label{fig:Figure9}
\end{figure}

Finally, we show that the piezo voltage induced strain not only allows to control the magnetic anisotropy, but also can be used to induce an irreversible switching of magnetization. The solid black curve in Fig.~\ref{fig:Figure9}(a) shows the transverse resistivity loop $\rho_{\rm trans}(\mu_0H)$ at 40~K and $V_{\rm p}=+200$~V, with the external magnetic field oriented along $\alpha=335^\circ$. As indicated by the black arrows, $\mu_0H$ was swept from $-300$~mT to $+300$~mT and back to $-300$~mT. The resistivity jump in the upsweep at $\mu_0H\approx+4$~mT is due to a switching of magnetization orientation between two minima in the free energy surface. In the free energy diagram for $\mu_0H\approx-5$~mT shown in Fig.~\ref{fig:Figure9}(b) this would correspond to a switching from the minimum at $\beta\approx90^\circ$ into the minimum at $\beta\approx 250^\circ$. The hysteresis of $\rho_{\rm trans}(\mu_0H)$ between up- and downsweeps is caused by the magnetization residing in one of these two minima. To induce a switching process of magnetization orientation back to $\beta\approx90^\circ$, a negative magnetic field in excess of the coercive field $\mu_0H_C=-6$~mT has to be applied. As we will show in the following, the additional degree of freedom offered by the piezoelectric actuator allows to induce a nonreversible magnetization switching already at external magnetic fields below the coercive field.

Starting from the same magnetic preparation ($\mu_0H=-300$~mT, $V_{\rm p}=+200$~V) the upsweep [blue symbols in Fig.~\ref{fig:Figure9}(a)] is stopped in a second experiment at $\mu_0H=-5$~mT (position denoted $B_1$). Keeping the magnetic field constant at this value, the piezo voltage is swept from $V_{\rm p}=+200$~V to $V_{\rm p}=-200$~V with the corresponding increase in $\rho_{\rm trans}$ shown in Fig.~\ref{fig:Figure9}(c) [$B_1$($V_{\rm p}=+200$~V)$\rightarrow B_2$($V_{\rm p}=0$~V)$ \rightarrow B_3$($V_{\rm p}=-200$~V)]. Sweeping the piezo voltage back to $V_{\rm p}=+200$~V, $\rho_{\rm trans}(\mu_0H)$ remains in this high resistance state [$B_3$($V_{\rm p}=-200$~V)$\rightarrow B_4$($V_{\rm p}=0$~V)$ \rightarrow B_5$($V_{\rm p}=+200$~V)]. This behavior can be understood with the help of the corresponding free energy plots at $V_{\rm p} \in$\{+200~V, 0~V, -200~V\} in Fig.~\ref{fig:Figure9}(b). These have been derived from ADMR measurements as described in Sec.~\ref{sec:limmer}. The free energy plots indicate that the magnetization orientation is rotated from $\beta\approx 90^\circ$ at $V_{\rm p}=+200$~V ($B_1$) over $\beta\approx 135^\circ$ at $V_{\rm p}=0$~V ($B_2$) to $\beta\approx +200^\circ$ at $V_{\rm p}=-200$~V ($B_3$). Sweeping $V_{\rm p}$ back to $+200$~V, due to the potential barrier at $\beta\approx 180^\circ$, the magnetization does not rotate back in the same way, but evolves into the potential minimum at $\beta\approx+250^\circ$ ($B_3\rightarrow B_4 \rightarrow B_5$). Therefore, the sweep in piezo voltage from $V_{\rm p}=+200$~V to $V_{\rm p}=-200$~V and back to $V_{\rm p}=+200$~V results in an irreversible magnetization orientation change and the corresponding resistivity change observed at $\mu_0H=-5$~mT. We note that these experimental results are closely comparable to those reported in manuscript~\cite{Rus08}.

Sweeping $\mu_0H$ from -5~mT to +6~mT ($B_5 \rightarrow C$) after the piezo-loop cycling of $V_{\rm p}$ from +200~V to -200~V and back to +200~V, $\rho_{\rm trans}$ decreases and approximately follows the resistivity curve obtained in the conventional magnetotransport measurement for the opposite sweep direction in this magnetic field range. The fact that $\rho_{\rm trans}$ does not exactly coincide with the downsweep curve we attribute to first indications of multidomain effects, which begin to become important at these small magnetic fields. In particular, the downsweep curve  depicted as the full line in Fig.~\ref{fig:Figure9}(a) was measured after saturating the magnetization at +300~mT, so that multi-domain effects should be less important. As we sweep $\mu_0H$ back to -5~mT ($C \rightarrow B_5$) and up to +2~mT again ($B_5 \rightarrow D$) $\rho_{\rm trans}$ remains on this resistivity branch.

At $\mu_0H=+2$~mT ($D$) we performed a second irreversible switching process via cycling the piezo-voltage $V_{\rm p}$ again from $+200$~V to $-200$~V and to +200~V. In analogy to the switching process at $-5$~mT described above, the magnetization is switched from the minimum at $\beta\approx250^\circ$ into the minimum at $\beta\approx 90^\circ$ ($D\rightarrow E$). Sweeping $\mu_0H$ to $+3$~mT ($F$) after this second irreversible piezo-induced switching and back to $-6$~mT ($G$), $\rho_{\rm trans}$ as expected follows the corresponding upsweep branch. These experiments clearly demonstrate that the piezo voltage not only allows to control the magnetic anisotropy and thus the magnetization orientation, but also to induce an irreversible magnetization switching.

\section{\label{sec:conclusion}CONCLUSIONS}

In summary, we have investigated the magnetic properties of a piezoelectric actuator/ferromagnet  hybrid structure. Application of a voltage to the piezoelectric actuator results in an additional uniaxial magnetoelastic contribution to magnetic anisotropy. For the GaMnAs thin film investigated, the axis of elongation (contraction) corresponds to the magnetic hard (easy) axis of this uniaxial magnetic anisotropy. We studied the temperature dependence of the magnetic anisotropy including the magnetoelastic contribution using anisotropic magnetoresistance techniques. The temperature dependence of the derived magnetostrictive constant $\lambda_{111}$ is in agreement with the earlier results of Masmanidis \textit{et al.}~\cite{Mas05}. At $T=5$~K the magnetoelastic term constitutes only a minor contribution compared to the dominating cubic term, but we showed that the switching fields of $\rho(\mu_0H)$ loops are shifted by the application of a piezo voltage at this temperature. At 50~K $-$ where the magnetoelastic term dominates magnetic anisotropy $-$ we demonstrated a continuous, fully reversible control of magnetization orientation by about $70^\circ$, solely via application of voltage to the piezoelectric actuator. This piezo-voltage control of magnetization orientation is directly transferable to other ferromagnetic/piezoelectric hybrid structures opening the way to new innovative multifunctional device concepts, such as all electrically controlled magnetic memory elements. As an example, the piezo-voltage induced irreversible magnetization switching demonstrated at $T=40$~K constitutes the basic principle of a nonvolatile memory element.

In this paper, we have demonstrated the manipulation of the magnetization in the plane of a dilute magnetic semiconductor. However, also a manipulation of the magnetization from in-plane to out-of-plane should be possible. For this, the magnetic anisotropy landscape has to be changed by reducing the strong uniaxial anisotropy in growth direction present in Ga$_{1-x}$Mn$_x$As thin films grown on GaAs. As discussed in Sec.~\ref{sec:introduction}, this can be achieved by growing on a lattice matched Ga$_{1-y}$In$_y$As virtual substrate \cite{Liu03, Liu2005, She97}. In the presence of a dominantly cubic magnetic anisotropy with only a small remaining uniaxial anisotropy in growth direction, the application of a biaxial tensile strain within the film plane would thus allow a switching of the magnetization orientation from in-plane to out-of-plane. Such controllable biaxial in-plane strain could be realized \textit{e.g.} via cementing the GaMnAs film onto the head side of the piezoelectric actuator. The controlled application of either uniaxial or biaxial stress would allow a full three dimensional control of the magnetization orientation.

\begin{figure}[tbp]
\includegraphics[width=0.6\textwidth]{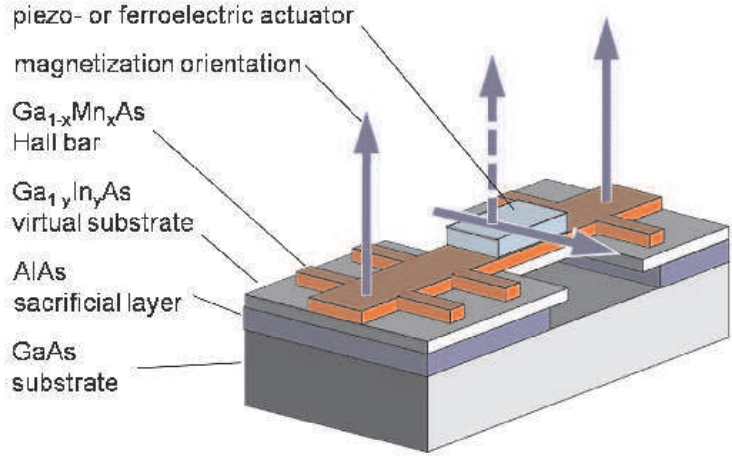} 
\caption{(color online) Sketch of a spin valve structure based on strain modulation. The magnetization of the central part of the Hall bar is manipulated by the actuator positioned above a Ga$_{1-x}$Mn$_x$As layer prestrained by a self-supporting Ga$_{1-y}$In$_y$As virtual substrate. In this particular realization, the Ga$_{1-y}$In$_y$As buffer with a lattice constant greater than that of the GaMnAs to layer leads to an out-of-plane orientation of the magnetization, which is flipped in-plane by a biaxial compressive strain induced by the actuator.}
\label{fig:device}
\end{figure}

A possible device concept based on this approach is exemplarily illustrated in Fig.~\ref{fig:device}, consisting of a GaMnAs layer which is strained locally by a piezo- or ferroelectric actuator. To allow optimal transfer of the stress from the actuator to the DMS layer, the Ga$_{1-x}$Mn$_x$As thin film should be made self-supporting by removing an AlAs sacrificial layer below it via wet-chemical etching. Finally, prestraining of the DMS can be achieved by the Ga$_{1-y}$In$_y$As virtual substrate discussed above or nanopatterning. To realize efficient strain transducers, a direct deposition of the actuator material onto the DMS is required. The fabrication of standard piezo- or ferroelectrics, such as PZT or BaTiO$_3$, is well established, but typically requires temperatures in excess of $400^\circ$C \cite{Mat94, Ale06}. Therefore, these deposition processes cannot be straightforwardly transferred to GaMnAs films, as this ferromagnetic semiconductor is grown by LT-MBE at substrate temperatures around $250^\circ$C or below \cite{Ohn01}. The exposure of GaMnAs to temperatures above $250^\circ$C results in Mn segregation and MnAs formation and thus to a destruction of the ferromagnetic semiconductor. To circumvent this issue, more exotic fabrication routes, e.g. the realization of PZT or BaTiO$_3$ via hydrothermal methods \cite{Mor04, Mor06}, could be viable approaches. Organic ferroelectrics such as poly(vinylidene fluoride/trifluoroethylene) or PVDF/TrFE are an interesting alternative, as they can be spin-coated and polymerized at temperatures below $200^\circ$C and therefore should be compatible with GaMnAs \cite{Fur89, Nab05}.

\section*{ACKNOWLEDGMENTS}
This work was supported by DFG through SPP 1157 GR 1132/13 (Walther-Meissner-Institut), LI 988/4 (Ulm), and SFB 631 (Walter Schottky Institut).

\section*{APPENDIX: MAGNETOELASTIC CONTRIBUTION TO MAGNETIC ANISOTROPY ENERGY}
In this appendix we discuss the influence of the piezo-voltage induced strain on the magnetic anisotropy of the film. As described in section~\ref{sec:experimental} the (001) oriented sample is cemented onto the piezoelectric actuator with the main expansion direction parallel to the [110]-direction of GaAs. In the $\left\{\textbf{j},\textbf{t},\textbf{n}\right\}$ coordinate system with $\textbf{j}||[110]$, $\textbf{t}||[\overline{1}10]$, and $\textbf{n}||[001]$ the strain induced by the piezoelectric actuator in the GaMnAs film is given by the tensor
\begin{equation}
\epsilon_{\left\{\textbf{j},\textbf{t},\textbf{n}\right\}}=
\left(
  \begin{array}{ccc}
    \epsilon_{jj} & 0 & 0 \\
    0 & \epsilon_{tt} & 0 \\
    0 & 0 & \epsilon_{nn} \\
  \end{array}
\right)
\end{equation}
with the diagonal components $\epsilon_{ii}$ describing the strains along the coordinate axes. Note that all other components vanish since in the $\left\{\textbf{j},\textbf{t},\textbf{n}\right\}$ coordinate system no shear strains are present. To derive the piezo-induced change in magnetic anisotropy, we transform $\epsilon_{\left\{\textbf{j},\textbf{t},\textbf{n}\right\}}$ into the $\left\{\textbf{x},\textbf{y},\textbf{z}\right\}$ coordinate system with coordinate axes along $\left<100\right>$ using the transformation matrix
\begin{equation}
T=
\left(
  \begin{array}{ccc}
    \sqrt{\frac{1}{2}} & \sqrt{\frac{1}{2}} & 0 \\
    -\sqrt{\frac{1}{2}} & \sqrt{\frac{1}{2}} & 0 \\
    0 & 0 & 1 \\
  \end{array}
\right)
\end{equation}
and its inverted matrix
\begin{equation}
T^{-1}=
\left(
  \begin{array}{ccc}
    \sqrt{\frac{1}{2}} & -\sqrt{\frac{1}{2}} & 0 \\
    \sqrt{\frac{1}{2}} & \sqrt{\frac{1}{2}} & 0 \\
    0 & 0 & 1 \\
  \end{array}
\right).
\end{equation}
In this way we obtain the strain tensor in the $\left\{\textbf{x},\textbf{y},\textbf{z}\right\}$ coordinate system
\begin{equation}
\epsilon_{\left\{\textbf{x},\textbf{y},\textbf{z}\right\}}=T\epsilon_{\left\{\textbf{j},\textbf{t},\textbf{n}\right\}}T^{-1}=
\left(
  \begin{array}{ccc}
    \frac{1}{2}(\epsilon_{jj}+\epsilon_{tt}) & -\frac{1}{2}(\epsilon_{jj}-\epsilon_{tt}) & 0 \\
    -\frac{1}{2}(\epsilon_{jj}-\epsilon_{tt}) & \frac{1}{2}(\epsilon_{jj}+\epsilon_{tt}) & 0 \\
    0 & 0 & \epsilon_{nn} \\
  \end{array}
\right)
\end{equation}
with the strains $\epsilon_{xx}=\epsilon_{yy}=\frac{1}{2}(\epsilon_{jj}+\epsilon_{tt})$ and $\epsilon_{zz}=\epsilon_{nn}$ and the shear strains $\epsilon_{xy}=\epsilon_{yx}=-\frac{1}{2}(\epsilon_{jj}-\epsilon_{tt})$. These are now inserted in the general expression for the magnetoelastic contribution to the magnetic anisotropy \cite{Chi64}
\begin{eqnarray}
F_{\rm magel}&=&B_1\left[\epsilon_{xx}\left(m_x^2-\frac{1}{3}\right)+\epsilon_{yy}\left(m_y^2-\frac{1}{3}\right)+\epsilon_{zz}\left(m_z^2-\frac{1}{3}\right)\right]  \nonumber \\
&+&B_2\left(\epsilon_{xy} m_x m_y+\epsilon_{yz} m_y m_z+\epsilon_{xz} m_z m_x\right)
\label{equ:FmagelChikazumi}
\end{eqnarray}
with $m_i$ denoting the direction cosines of magnetization orientation with respect to the cubic axes, the magnetoelastic coupling constants  $B_i$, and $\epsilon_{ij}$ describing the strains in the $\left\{\textbf{x},\textbf{y},\textbf{z}\right\}$ coordinate system. The magnetoelastic coupling constants $B_i$ can be substituted with the magnetostrictive constants $\lambda_{100}$ and $\lambda_{111}$ using the relations
$B_1=\frac{3}{2}\lambda_{100}(c_{12}-c_{11})$ and $B_2=-3\lambda_{111}c_{44}$ with the elastic moduli $c_{ij}$ \cite{Chi64}.

Since $\epsilon_{xx}=\epsilon_{yy}$, $m_x^2+m_y^2=1$, and magnetization is oriented in the film plane ($m_z=0$), the first term in (\ref{equ:FmagelChikazumi}) exhibits no anisotropy and therefore in the following is omitted. Furthermore, using $m_x m_y = \cos(\beta+45^\circ) \cos(\beta-45^\circ) = \cos^2(\beta)-\frac{1}{2}$ and again omitting the isotropic part we obtain
\begin{equation}
F_{\rm magel}/M=\frac{3\lambda_{111}c_{44}(\epsilon_{jj}-\epsilon_{tt})}{2M}\cos^2\beta=B_{110}m^2_j.
\label{equ:FmagelChikazumi2}
\end{equation}

This expression consitutes a uniaxial magnetic anisotropy contribution along [110] with the direction cosine of magnetization orientation relative to the \textbf{j}-axis $m_j$ and the corresponding uniaxial anisotropy parameter $B_{110}=\frac{3\lambda_{111}c_{44}(\epsilon_{jj}-\epsilon_{tt})}{2M}$ and is used as the third term in the total free energy in Eq.~(\ref{eq:F}).


\begin{thebibliography}{66}
\expandafter\ifx\csname natexlab\endcsname\relax\def\natexlab#1{#1}\fi
\expandafter\ifx\csname bibnamefont\endcsname\relax
  \def\bibnamefont#1{#1}\fi
\expandafter\ifx\csname bibfnamefont\endcsname\relax
  \def\bibfnamefont#1{#1}\fi
\expandafter\ifx\csname citenamefont\endcsname\relax
  \def\citenamefont#1{#1}\fi
\expandafter\ifx\csname url\endcsname\relax
  \def\url#1{\texttt{#1}}\fi
\expandafter\ifx\csname urlprefix\endcsname\relax\def\urlprefix{URL }\fi
\providecommand{\bibinfo}[2]{#2}
\providecommand{\eprint}[2][]{\url{#2}}

\bibitem[{\citenamefont{Ohno}(1998)}]{Ohn98}
\bibinfo{author}{\bibfnamefont{H.}~\bibnamefont{Ohno}},
  \bibinfo{journal}{Science} \textbf{\bibinfo{volume}{281}},
  \bibinfo{pages}{951} (\bibinfo{year}{1998}).

\bibitem[{\citenamefont{Dietl et~al.}(2000)\citenamefont{Dietl, Ohno,
  Matsukura, Cibert, and Ferrand}}]{Die00}
\bibinfo{author}{\bibfnamefont{T.}~\bibnamefont{Dietl}},
  \bibinfo{author}{\bibfnamefont{H.}~\bibnamefont{Ohno}},
  \bibinfo{author}{\bibfnamefont{F.}~\bibnamefont{Matsukura}},
  \bibinfo{author}{\bibfnamefont{J.}~\bibnamefont{Cibert}}, \bibnamefont{and}
  \bibinfo{author}{\bibfnamefont{D.}~\bibnamefont{Ferrand}},
  \bibinfo{journal}{Science} \textbf{\bibinfo{volume}{287}},
  \bibinfo{pages}{1019} (\bibinfo{year}{2000}).

\bibitem[{\citenamefont{Wolf et~al.}(2001)\citenamefont{Wolf, Awschalom,
  Buhrman, Daughton, Moln$\acute{\rm a}$r, Roukes, Chtchelkanova, and
  Treger}}]{Wol01}
\bibinfo{author}{\bibfnamefont{S.~A.} \bibnamefont{Wolf}},
  \bibinfo{author}{\bibfnamefont{D.~D.} \bibnamefont{Awschalom}},
  \bibinfo{author}{\bibfnamefont{R.~A.} \bibnamefont{Buhrman}},
  \bibinfo{author}{\bibfnamefont{J.~M.} \bibnamefont{Daughton}},
  \bibinfo{author}{\bibfnamefont{S.~v.} \bibnamefont{Moln$\acute{\rm a}$r}},
  \bibinfo{author}{\bibfnamefont{M.~L.} \bibnamefont{Roukes}},
  \bibinfo{author}{\bibfnamefont{A.~Y.} \bibnamefont{Chtchelkanova}},
  \bibnamefont{and} \bibinfo{author}{\bibfnamefont{D.~M.}
  \bibnamefont{Treger}}, \bibinfo{journal}{Science}
  \textbf{\bibinfo{volume}{294}}, \bibinfo{pages}{1488} (\bibinfo{year}{2001}).

\bibitem[{\citenamefont{R\"{u}ster et~al.}(2003)\citenamefont{R\"{u}ster,
  Borzenko, Gould, Schmidt, Molenkamp, Liu, Wojtowicz, Furdyna, Yu, and
  Flatt$\acute{\rm e}$}}]{Rue03}
\bibinfo{author}{\bibfnamefont{C.}~\bibnamefont{R\"{u}ster}},
  \bibinfo{author}{\bibfnamefont{T.}~\bibnamefont{Borzenko}},
  \bibinfo{author}{\bibfnamefont{C.}~\bibnamefont{Gould}},
  \bibinfo{author}{\bibfnamefont{G.}~\bibnamefont{Schmidt}},
  \bibinfo{author}{\bibfnamefont{L.~W.} \bibnamefont{Molenkamp}},
  \bibinfo{author}{\bibfnamefont{X.}~\bibnamefont{Liu}},
  \bibinfo{author}{\bibfnamefont{T.~J.} \bibnamefont{Wojtowicz}},
  \bibinfo{author}{\bibfnamefont{J.~K.} \bibnamefont{Furdyna}},
  \bibinfo{author}{\bibfnamefont{Z.~G.} \bibnamefont{Yu}}, \bibnamefont{and}
  \bibinfo{author}{\bibfnamefont{M.~E.} \bibnamefont{Flatt$\acute{\rm e}$}},
  \bibinfo{journal}{Phys. Rev. Lett.} \textbf{\bibinfo{volume}{91}},
  \bibinfo{pages}{216602} (\bibinfo{year}{2003}).

\bibitem[{\citenamefont{Pearton et~al.}(2005)\citenamefont{Pearton, Norton,
  Frazier, Han, Abernathy, and Zavada}}]{Pea05}
\bibinfo{author}{\bibfnamefont{S.~J.} \bibnamefont{Pearton}},
  \bibinfo{author}{\bibfnamefont{D.~P.} \bibnamefont{Norton}},
  \bibinfo{author}{\bibfnamefont{R.}~\bibnamefont{Frazier}},
  \bibinfo{author}{\bibfnamefont{S.~Y.} \bibnamefont{Han}},
  \bibinfo{author}{\bibfnamefont{C.~R.} \bibnamefont{Abernathy}},
  \bibnamefont{and} \bibinfo{author}{\bibfnamefont{J.~M.}
  \bibnamefont{Zavada}}, \bibinfo{journal}{IEEE Proc.-Circuits Devices Syst.}
  \textbf{\bibinfo{volume}{152}}, \bibinfo{pages}{312} (\bibinfo{year}{2005}).

\bibitem[{\citenamefont{Jungwirth et~al.}(2006)\citenamefont{Jungwirth, Sinova,
  Ma$\check{\rm s}$ek, Ku$\check{\rm c}$era, and MacDonald}}]{Jun2006}
\bibinfo{author}{\bibfnamefont{T.}~\bibnamefont{Jungwirth}},
  \bibinfo{author}{\bibfnamefont{J.}~\bibnamefont{Sinova}},
  \bibinfo{author}{\bibfnamefont{J.}~\bibnamefont{Ma$\check{\rm s}$ek}},
  \bibinfo{author}{\bibfnamefont{J.}~\bibnamefont{Ku$\check{\rm c}$era}},
  \bibnamefont{and} \bibinfo{author}{\bibfnamefont{A.~H.}
  \bibnamefont{MacDonald}}, \bibinfo{journal}{Rev. Mod. Phys.}
  \textbf{\bibinfo{volume}{78}}, \bibinfo{pages}{809} (\bibinfo{year}{2006}).

\bibitem[{\citenamefont{Figielski et~al.}(2007)\citenamefont{Figielski,
  Wosinski, Morawski, Makosa, Wrobel, and Sadowski}}]{Fig07}
\bibinfo{author}{\bibfnamefont{T.}~\bibnamefont{Figielski}},
  \bibinfo{author}{\bibfnamefont{T.}~\bibnamefont{Wosinski}},
  \bibinfo{author}{\bibfnamefont{A.}~\bibnamefont{Morawski}},
  \bibinfo{author}{\bibfnamefont{A.}~\bibnamefont{Makosa}},
  \bibinfo{author}{\bibfnamefont{J.}~\bibnamefont{Wrobel}}, \bibnamefont{and}
  \bibinfo{author}{\bibfnamefont{J.}~\bibnamefont{Sadowski}},
  \bibinfo{journal}{Appl. Phys. Lett.} \textbf{\bibinfo{volume}{90}},
  \bibinfo{pages}{052108} (\bibinfo{year}{2007}).

\bibitem[{\citenamefont{Pappert et~al.}(2007)\citenamefont{Pappert,
  H\"{u}mpfner, Gould, Wenisch, Brunner, Schmidt, and Molenkamp}}]{Pap07}
\bibinfo{author}{\bibfnamefont{K.}~\bibnamefont{Pappert}},
  \bibinfo{author}{\bibfnamefont{S.}~\bibnamefont{H\"{u}mpfner}},
  \bibinfo{author}{\bibfnamefont{C.}~\bibnamefont{Gould}},
  \bibinfo{author}{\bibfnamefont{J.}~\bibnamefont{Wenisch}},
  \bibinfo{author}{\bibfnamefont{K.}~\bibnamefont{Brunner}},
  \bibinfo{author}{\bibfnamefont{G.}~\bibnamefont{Schmidt}}, \bibnamefont{and}
  \bibinfo{author}{\bibfnamefont{L.~W.} \bibnamefont{Molenkamp}},
  \bibinfo{journal}{Nature Phys.} \textbf{\bibinfo{volume}{3}},
  \bibinfo{pages}{573} (\bibinfo{year}{2007}).

\bibitem[{\citenamefont{Dietl et~al.}(2001)\citenamefont{Dietl, Ohno, and
  Matsukura}}]{Die01}
\bibinfo{author}{\bibfnamefont{T.}~\bibnamefont{Dietl}},
  \bibinfo{author}{\bibfnamefont{H.}~\bibnamefont{Ohno}}, \bibnamefont{and}
  \bibinfo{author}{\bibfnamefont{F.}~\bibnamefont{Matsukura}},
  \bibinfo{journal}{Phys. Rev. B} \textbf{\bibinfo{volume}{63}},
  \bibinfo{pages}{195205} (\bibinfo{year}{2001}).

\bibitem[{\citenamefont{Tang et~al.}(2003)\citenamefont{Tang, Kawakami,
  Awschalom, and Roukes}}]{Tan03}
\bibinfo{author}{\bibfnamefont{H.~X.} \bibnamefont{Tang}},
  \bibinfo{author}{\bibfnamefont{R.~K.} \bibnamefont{Kawakami}},
  \bibinfo{author}{\bibfnamefont{D.~D.} \bibnamefont{Awschalom}},
  \bibnamefont{and} \bibinfo{author}{\bibfnamefont{M.~L.}
  \bibnamefont{Roukes}}, \bibinfo{journal}{Phys. Rev. Lett.}
  \textbf{\bibinfo{volume}{90}}, \bibinfo{pages}{107201}
  (\bibinfo{year}{2003}).

\bibitem[{\citenamefont{Ohno and Matsukura}(2001)}]{Ohn01}
\bibinfo{author}{\bibfnamefont{H.}~\bibnamefont{Ohno}} \bibnamefont{and}
  \bibinfo{author}{\bibfnamefont{F.}~\bibnamefont{Matsukura}},
  \bibinfo{journal}{Solid State Commun.} \textbf{\bibinfo{volume}{117}},
  \bibinfo{pages}{179} (\bibinfo{year}{2001}).

\bibitem[{\citenamefont{Van~Dorpe et~al.}(2004)\citenamefont{Van~Dorpe, Liu,
  Van~Roy, Motsnyi, Sawicki, Borghs, and De~Boeck}}]{Dor04}
\bibinfo{author}{\bibfnamefont{P.}~\bibnamefont{Van~Dorpe}},
  \bibinfo{author}{\bibfnamefont{Z.}~\bibnamefont{Liu}},
  \bibinfo{author}{\bibfnamefont{W.}~\bibnamefont{Van~Roy}},
  \bibinfo{author}{\bibfnamefont{V.~F.} \bibnamefont{Motsnyi}},
  \bibinfo{author}{\bibfnamefont{M.}~\bibnamefont{Sawicki}},
  \bibinfo{author}{\bibfnamefont{G.}~\bibnamefont{Borghs}}, \bibnamefont{and}
  \bibinfo{author}{\bibfnamefont{J.}~\bibnamefont{De~Boeck}},
  \bibinfo{journal}{Appl. Phys. Lett.} \textbf{\bibinfo{volume}{84}},
  \bibinfo{pages}{3495} (\bibinfo{year}{2004}).

\bibitem[{\citenamefont{Braden et~al.}(2003)\citenamefont{Braden, Parker,
  Xiong, Chun, and Samarth}}]{Bra03}
\bibinfo{author}{\bibfnamefont{J.~G.} \bibnamefont{Braden}},
  \bibinfo{author}{\bibfnamefont{J.~S.} \bibnamefont{Parker}},
  \bibinfo{author}{\bibfnamefont{P.}~\bibnamefont{Xiong}},
  \bibinfo{author}{\bibfnamefont{S.~H.} \bibnamefont{Chun}}, \bibnamefont{and}
  \bibinfo{author}{\bibfnamefont{N.}~\bibnamefont{Samarth}},
  \bibinfo{journal}{Phys. Rev. Lett.} \textbf{\bibinfo{volume}{91}},
  \bibinfo{pages}{056602} (\bibinfo{year}{2003}).

\bibitem[{\citenamefont{Tanaka and Higo}(2001)}]{Tan01}
\bibinfo{author}{\bibfnamefont{M.}~\bibnamefont{Tanaka}} \bibnamefont{and}
  \bibinfo{author}{\bibfnamefont{Y.}~\bibnamefont{Higo}},
  \bibinfo{journal}{Phys. Rev. Lett.} \textbf{\bibinfo{volume}{87}},
  \bibinfo{pages}{026602} (\bibinfo{year}{2001}).

\bibitem[{\citenamefont{Welp et~al.}(2003)\citenamefont{Welp, Vlasko-Vlasov,
  Liu, Furdyna, and Wojtowicz}}]{Wel03}
\bibinfo{author}{\bibfnamefont{U.}~\bibnamefont{Welp}},
  \bibinfo{author}{\bibfnamefont{V.~K.} \bibnamefont{Vlasko-Vlasov}},
  \bibinfo{author}{\bibfnamefont{X.}~\bibnamefont{Liu}},
  \bibinfo{author}{\bibfnamefont{J.~K.} \bibnamefont{Furdyna}},
  \bibnamefont{and}
  \bibinfo{author}{\bibfnamefont{T.}~\bibnamefont{Wojtowicz}},
  \bibinfo{journal}{Phys. Rev . Lett.} \textbf{\bibinfo{volume}{90}},
  \bibinfo{pages}{167206} (\bibinfo{year}{2003}).

\bibitem[{\citenamefont{Ohno et~al.}(1999)\citenamefont{Ohno, Young, Beschoten,
  Matsukura, Ohno, and Awschalom}}]{Ohn99}
\bibinfo{author}{\bibfnamefont{Y.}~\bibnamefont{Ohno}},
  \bibinfo{author}{\bibfnamefont{D.~K.} \bibnamefont{Young}},
  \bibinfo{author}{\bibfnamefont{B.}~\bibnamefont{Beschoten}},
  \bibinfo{author}{\bibfnamefont{F.}~\bibnamefont{Matsukura}},
  \bibinfo{author}{\bibfnamefont{H.}~\bibnamefont{Ohno}}, \bibnamefont{and}
  \bibinfo{author}{\bibfnamefont{D.~D.} \bibnamefont{Awschalom}},
  \bibinfo{journal}{Nature} \textbf{\bibinfo{volume}{402}},
  \bibinfo{pages}{790} (\bibinfo{year}{1999}).

\bibitem[{\citenamefont{Yamanouchi et~al.}(2004)\citenamefont{Yamanouchi,
  Chiba, Matsukura, and Ohno}}]{Yam04}
\bibinfo{author}{\bibfnamefont{M.}~\bibnamefont{Yamanouchi}},
  \bibinfo{author}{\bibfnamefont{D.}~\bibnamefont{Chiba}},
  \bibinfo{author}{\bibfnamefont{F.}~\bibnamefont{Matsukura}},
  \bibnamefont{and} \bibinfo{author}{\bibfnamefont{H.}~\bibnamefont{Ohno}},
  \bibinfo{journal}{Nature} \textbf{\bibinfo{volume}{428}},
  \bibinfo{pages}{539} (\bibinfo{year}{2004}).

\bibitem[{\citenamefont{Yamanouchi et~al.}(2006)\citenamefont{Yamanouchi,
  Chiba, Matsukura, Dietl, and Ohno}}]{Yam06}
\bibinfo{author}{\bibfnamefont{M.}~\bibnamefont{Yamanouchi}},
  \bibinfo{author}{\bibfnamefont{D.}~\bibnamefont{Chiba}},
  \bibinfo{author}{\bibfnamefont{F.}~\bibnamefont{Matsukura}},
  \bibinfo{author}{\bibfnamefont{T.}~\bibnamefont{Dietl}}, \bibnamefont{and}
  \bibinfo{author}{\bibfnamefont{H.}~\bibnamefont{Ohno}},
  \bibinfo{journal}{Phys. Rev. Lett.} \textbf{\bibinfo{volume}{96}},
  \bibinfo{pages}{096601} (\bibinfo{year}{2006}).

\bibitem[{\citenamefont{Gould et~al.}(2004)\citenamefont{Gould, R\"{u}ster,
  Jungwirth, Girgis, Schott, Giraud, Brunner, Schmidt, and Molenkamp}}]{Gou04}
\bibinfo{author}{\bibfnamefont{C.}~\bibnamefont{Gould}},
  \bibinfo{author}{\bibfnamefont{C.}~\bibnamefont{R\"{u}ster}},
  \bibinfo{author}{\bibfnamefont{T.}~\bibnamefont{Jungwirth}},
  \bibinfo{author}{\bibfnamefont{E.}~\bibnamefont{Girgis}},
  \bibinfo{author}{\bibfnamefont{G.~M.} \bibnamefont{Schott}},
  \bibinfo{author}{\bibfnamefont{R.}~\bibnamefont{Giraud}},
  \bibinfo{author}{\bibfnamefont{K.}~\bibnamefont{Brunner}},
  \bibinfo{author}{\bibfnamefont{G.}~\bibnamefont{Schmidt}}, \bibnamefont{and}
  \bibinfo{author}{\bibfnamefont{L.~W.} \bibnamefont{Molenkamp}},
  \bibinfo{journal}{Phys. Rev. Lett.} \textbf{\bibinfo{volume}{93}},
  \bibinfo{pages}{117203} (\bibinfo{year}{2004}).

\bibitem[{\citenamefont{Liu et~al.}(2003)\citenamefont{Liu, Sasaki, and
  Furdyna}}]{Liu03}
\bibinfo{author}{\bibfnamefont{X.}~\bibnamefont{Liu}},
  \bibinfo{author}{\bibfnamefont{Y.}~\bibnamefont{Sasaki}}, \bibnamefont{and}
  \bibinfo{author}{\bibfnamefont{J.~K.} \bibnamefont{Furdyna}},
  \bibinfo{journal}{Phys. Rev. B} \textbf{\bibinfo{volume}{67}},
  \bibinfo{pages}{205204} (\bibinfo{year}{2003}).

\bibitem[{\citenamefont{Liu et~al.}(2005)\citenamefont{Liu, Lim, Titova,
  Dobrowolska, Furdyna, Kutrowski, and Wojtowicz}}]{Liu2005}
\bibinfo{author}{\bibfnamefont{X.}~\bibnamefont{Liu}},
  \bibinfo{author}{\bibfnamefont{W.~L.} \bibnamefont{Lim}},
  \bibinfo{author}{\bibfnamefont{L.~V.} \bibnamefont{Titova}},
  \bibinfo{author}{\bibfnamefont{M.}~\bibnamefont{Dobrowolska}},
  \bibinfo{author}{\bibfnamefont{J.~K.} \bibnamefont{Furdyna}},
  \bibinfo{author}{\bibfnamefont{M.}~\bibnamefont{Kutrowski}},
  \bibnamefont{and}
  \bibinfo{author}{\bibfnamefont{T.}~\bibnamefont{Wojtowicz}},
  \bibinfo{journal}{J. Appl. Phys.} \textbf{\bibinfo{volume}{98}},
  \bibinfo{pages}{063904} (\bibinfo{year}{2005}).

\bibitem[{\citenamefont{Shen et~al.}(1997)\citenamefont{Shen, Ohno, Matsukura,
  Sugawara, Akiba, Kuroiwa, Oiwa, Endo, Katsumota, and Iye}}]{She97}
\bibinfo{author}{\bibfnamefont{A.}~\bibnamefont{Shen}},
  \bibinfo{author}{\bibfnamefont{H.}~\bibnamefont{Ohno}},
  \bibinfo{author}{\bibfnamefont{F.}~\bibnamefont{Matsukura}},
  \bibinfo{author}{\bibfnamefont{Y.}~\bibnamefont{Sugawara}},
  \bibinfo{author}{\bibfnamefont{N.}~\bibnamefont{Akiba}},
  \bibinfo{author}{\bibfnamefont{T.}~\bibnamefont{Kuroiwa}},
  \bibinfo{author}{\bibfnamefont{A.}~\bibnamefont{Oiwa}},
  \bibinfo{author}{\bibfnamefont{A.}~\bibnamefont{Endo}},
  \bibinfo{author}{\bibfnamefont{S.}~\bibnamefont{Katsumota}},
  \bibnamefont{and} \bibinfo{author}{\bibfnamefont{Y.}~\bibnamefont{Iye}},
  \bibinfo{journal}{J. Cryst. Growth} \textbf{\bibinfo{volume}{175/176}},
  \bibinfo{pages}{1069} (\bibinfo{year}{1997}).

\bibitem[{\citenamefont{Sawicki et~al.}(2004)\citenamefont{Sawicki, Matsukura,
  Idziaszek, Dietl, Schott, Ruester, Gould, Karczewski, Schmidt, and
  Molenkamp}}]{Saw04}
\bibinfo{author}{\bibfnamefont{M.}~\bibnamefont{Sawicki}},
  \bibinfo{author}{\bibfnamefont{F.}~\bibnamefont{Matsukura}},
  \bibinfo{author}{\bibfnamefont{A.}~\bibnamefont{Idziaszek}},
  \bibinfo{author}{\bibfnamefont{T.}~\bibnamefont{Dietl}},
  \bibinfo{author}{\bibfnamefont{G.~M.} \bibnamefont{Schott}},
  \bibinfo{author}{\bibfnamefont{C.}~\bibnamefont{Ruester}},
  \bibinfo{author}{\bibfnamefont{C.}~\bibnamefont{Gould}},
  \bibinfo{author}{\bibfnamefont{G.}~\bibnamefont{Karczewski}},
  \bibinfo{author}{\bibfnamefont{G.}~\bibnamefont{Schmidt}}, \bibnamefont{and}
  \bibinfo{author}{\bibfnamefont{L.~W.} \bibnamefont{Molenkamp}},
  \bibinfo{journal}{Phys. Rev. B} \textbf{\bibinfo{volume}{70}},
  \bibinfo{pages}{245325} (\bibinfo{year}{2004}).

\bibitem[{\citenamefont{Hamaya et~al.}(2003)\citenamefont{Hamaya, Taniyama,
  Kitamoto, Moriya, and Munekata}}]{Ham03}
\bibinfo{author}{\bibfnamefont{K.}~\bibnamefont{Hamaya}},
  \bibinfo{author}{\bibfnamefont{T.}~\bibnamefont{Taniyama}},
  \bibinfo{author}{\bibfnamefont{Y.}~\bibnamefont{Kitamoto}},
  \bibinfo{author}{\bibfnamefont{R.}~\bibnamefont{Moriya}}, \bibnamefont{and}
  \bibinfo{author}{\bibfnamefont{H.}~\bibnamefont{Munekata}},
  \bibinfo{journal}{J. Appl. Phys.} \textbf{\bibinfo{volume}{94}},
  \bibinfo{pages}{7657} (\bibinfo{year}{2003}).

\bibitem[{\citenamefont{Welp et~al.}(2004)\citenamefont{Welp, Vlasko-Vlasov,
  Menzel, You, Liu, Furdyna, and Wojtowicz}}]{Wel04}
\bibinfo{author}{\bibfnamefont{U.}~\bibnamefont{Welp}},
  \bibinfo{author}{\bibfnamefont{V.~K.} \bibnamefont{Vlasko-Vlasov}},
  \bibinfo{author}{\bibfnamefont{A.}~\bibnamefont{Menzel}},
  \bibinfo{author}{\bibfnamefont{H.~D.} \bibnamefont{You}},
  \bibinfo{author}{\bibfnamefont{X.}~\bibnamefont{Liu}},
  \bibinfo{author}{\bibfnamefont{J.~K.} \bibnamefont{Furdyna}},
  \bibnamefont{and}
  \bibinfo{author}{\bibfnamefont{T.}~\bibnamefont{Wojtowicz}},
  \bibinfo{journal}{Appl. Phys. Lett.} \textbf{\bibinfo{volume}{85}},
  \bibinfo{pages}{260} (\bibinfo{year}{2004}).

\bibitem[{\citenamefont{Sawicki et~al.}(2005)\citenamefont{Sawicki, Wang,
  Edmonds, Campion, Staddon, Farley, Foxon, Papis, Kami$\acute{\rm n}$ska,
  Piotrowska et~al.}}]{Saw05}
\bibinfo{author}{\bibfnamefont{M.}~\bibnamefont{Sawicki}},
  \bibinfo{author}{\bibfnamefont{K.-Y.} \bibnamefont{Wang}},
  \bibinfo{author}{\bibfnamefont{K.~W.} \bibnamefont{Edmonds}},
  \bibinfo{author}{\bibfnamefont{R.~P.} \bibnamefont{Campion}},
  \bibinfo{author}{\bibfnamefont{C.~R.} \bibnamefont{Staddon}},
  \bibinfo{author}{\bibfnamefont{N.~R.~S.} \bibnamefont{Farley}},
  \bibinfo{author}{\bibfnamefont{C.~T.} \bibnamefont{Foxon}},
  \bibinfo{author}{\bibfnamefont{E.}~\bibnamefont{Papis}},
  \bibinfo{author}{\bibfnamefont{E.}~\bibnamefont{Kami$\acute{\rm n}$ska}},
  \bibinfo{author}{\bibfnamefont{A.}~\bibnamefont{Piotrowska}},
  \bibnamefont{et~al.}, \bibinfo{journal}{Phys. Rev. B}
  \textbf{\bibinfo{volume}{71}}, \bibinfo{pages}{121302(R)}
  (\bibinfo{year}{2005}).

\bibitem[{\citenamefont{Stanciu and Svedlindh}(2005)}]{Sta05}
\bibinfo{author}{\bibfnamefont{V.}~\bibnamefont{Stanciu}} \bibnamefont{and}
  \bibinfo{author}{\bibfnamefont{P.}~\bibnamefont{Svedlindh}},
  \bibinfo{journal}{Appl. Phys. Lett.} \textbf{\bibinfo{volume}{87}},
  \bibinfo{pages}{242509} (\bibinfo{year}{2005}).

\bibitem[{\citenamefont{Hamaya et~al.}(2005)\citenamefont{Hamaya, Taniyama,
  Kitamoto, Fuji, and Yamazaki}}]{Ham05}
\bibinfo{author}{\bibfnamefont{K.}~\bibnamefont{Hamaya}},
  \bibinfo{author}{\bibfnamefont{T.}~\bibnamefont{Taniyama}},
  \bibinfo{author}{\bibfnamefont{Y.}~\bibnamefont{Kitamoto}},
  \bibinfo{author}{\bibfnamefont{T.}~\bibnamefont{Fujii}}, \bibnamefont{and}
  \bibinfo{author}{\bibfnamefont{Y.}~\bibnamefont{Yamazaki}},
  \bibinfo{journal}{Phys. Rev. Lett.} \textbf{\bibinfo{volume}{94}},
  \bibinfo{pages}{147203} (\bibinfo{year}{2005}).

\bibitem[{\citenamefont{Wang et~al.}(2005)\citenamefont{Wang, Edmonds, Zhao,
  Sawicki, Campion, Gallagher, and Foxon}}]{Wan05}
\bibinfo{author}{\bibfnamefont{K.~Y.} \bibnamefont{Wang}},
  \bibinfo{author}{\bibfnamefont{K.~W.} \bibnamefont{Edmonds}},
  \bibinfo{author}{\bibfnamefont{L.~X.} \bibnamefont{Zhao}},
  \bibinfo{author}{\bibfnamefont{M.}~\bibnamefont{Sawicki}},
  \bibinfo{author}{\bibfnamefont{R.~P.} \bibnamefont{Campion}},
  \bibinfo{author}{\bibfnamefont{B.~L.} \bibnamefont{Gallagher}},
  \bibnamefont{and} \bibinfo{author}{\bibfnamefont{C.~T.} \bibnamefont{Foxon}},
  \bibinfo{journal}{Phys. Rev. B} \textbf{\bibinfo{volume}{72}},
  \bibinfo{pages}{115207} (\bibinfo{year}{2005}).

\bibitem[{\citenamefont{Hamaya et~al.}(2006{\natexlab{a}})\citenamefont{Hamaya,
  Koike, Taniyama, Fuji, Kitamoto, and Yamazaki}}]{Ham06a}
\bibinfo{author}{\bibfnamefont{K.}~\bibnamefont{Hamaya}},
  \bibinfo{author}{\bibfnamefont{T.}~\bibnamefont{Koike}},
  \bibinfo{author}{\bibfnamefont{T.}~\bibnamefont{Taniyama}},
  \bibinfo{author}{\bibfnamefont{T.}~\bibnamefont{Fujii}},
  \bibinfo{author}{\bibfnamefont{Y.}~\bibnamefont{Kitamoto}}, \bibnamefont{and}
  \bibinfo{author}{\bibfnamefont{Y.}~\bibnamefont{Yamazaki}},
  \bibinfo{journal}{Phys. Rev. B} \textbf{\bibinfo{volume}{73}},
  \bibinfo{pages}{155204} (\bibinfo{year}{2006}{\natexlab{a}}).

\bibitem[{\citenamefont{Hamaya et~al.}(2006{\natexlab{b}})\citenamefont{Hamaya,
  Watanabe, Taniyama, Oiwa, Kitamoto, and Yamazaki}}]{Ham06b}
\bibinfo{author}{\bibfnamefont{K.}~\bibnamefont{Hamaya}},
  \bibinfo{author}{\bibfnamefont{T.}~\bibnamefont{Watanabe}},
  \bibinfo{author}{\bibfnamefont{T.}~\bibnamefont{Taniyama}},
  \bibinfo{author}{\bibfnamefont{A.}~\bibnamefont{Oiwa}},
  \bibinfo{author}{\bibfnamefont{Y.}~\bibnamefont{Kitamoto}}, \bibnamefont{and}
  \bibinfo{author}{\bibfnamefont{Y.}~\bibnamefont{Yamazaki}},
  \bibinfo{journal}{Phys. Rev. B} \textbf{\bibinfo{volume}{74}},
  \bibinfo{pages}{045201} (\bibinfo{year}{2006}{\natexlab{b}}).

\bibitem[{\citenamefont{Ohno et~al.}(2000)\citenamefont{Ohno, Chiba, Matsukura,
  Omiya, Abe, Dietl, Ohno, and Ohtani}}]{Ohn00}
\bibinfo{author}{\bibfnamefont{H.}~\bibnamefont{Ohno}},
  \bibinfo{author}{\bibfnamefont{D.}~\bibnamefont{Chiba}},
  \bibinfo{author}{\bibfnamefont{F.}~\bibnamefont{Matsukura}},
  \bibinfo{author}{\bibfnamefont{T.}~\bibnamefont{Omiya}},
  \bibinfo{author}{\bibfnamefont{E.}~\bibnamefont{Abe}},
  \bibinfo{author}{\bibfnamefont{T.}~\bibnamefont{Dietl}},
  \bibinfo{author}{\bibfnamefont{Y.}~\bibnamefont{Ohno}}, \bibnamefont{and}
  \bibinfo{author}{\bibfnamefont{K.}~\bibnamefont{Ohtani}},
  \bibinfo{journal}{Nature} \textbf{\bibinfo{volume}{408}},
  \bibinfo{pages}{944} (\bibinfo{year}{2000}).

\bibitem[{\citenamefont{Chiba et~al.}(2003)\citenamefont{Chiba, Yamanouchi,
  Matsukura, and Ohno}}]{Chi03a}
\bibinfo{author}{\bibfnamefont{D.}~\bibnamefont{Chiba}},
  \bibinfo{author}{\bibfnamefont{M.}~\bibnamefont{Yamanouchi}},
  \bibinfo{author}{\bibfnamefont{F.}~\bibnamefont{Matsukura}},
  \bibnamefont{and} \bibinfo{author}{\bibfnamefont{H.}~\bibnamefont{Ohno}},
  \bibinfo{journal}{Science} \textbf{\bibinfo{volume}{301}},
  \bibinfo{pages}{943} (\bibinfo{year}{2003}).

\bibitem[{\citenamefont{Koshihara et~al.}(1997)\citenamefont{Koshihara, Oiwa,
  Hirasawa, Katsumoto, Iye, Urano, Takagi, and Munekata}}]{Kos97}
\bibinfo{author}{\bibfnamefont{S.}~\bibnamefont{Koshihara}},
  \bibinfo{author}{\bibfnamefont{A.}~\bibnamefont{Oiwa}},
  \bibinfo{author}{\bibfnamefont{M.}~\bibnamefont{Hirasawa}},
  \bibinfo{author}{\bibfnamefont{S.}~\bibnamefont{Katsumoto}},
  \bibinfo{author}{\bibfnamefont{Y.}~\bibnamefont{Iye}},
  \bibinfo{author}{\bibfnamefont{C.}~\bibnamefont{Urano}},
  \bibinfo{author}{\bibfnamefont{H.}~\bibnamefont{Takagi}}, \bibnamefont{and}
  \bibinfo{author}{\bibfnamefont{H.}~\bibnamefont{Munekata}},
  \bibinfo{journal}{Phys. Rev. Lett.} \textbf{\bibinfo{volume}{78}},
  \bibinfo{pages}{4617} (\bibinfo{year}{1997}).

\bibitem[{\citenamefont{Boukari et~al.}(2002)\citenamefont{Boukari, Kossacki,
  Bertolini, Ferrand, Cibert, Tatarenko, Wasiela, Gaj, and Dietl}}]{Bou02}
\bibinfo{author}{\bibfnamefont{H.}~\bibnamefont{Boukari}},
  \bibinfo{author}{\bibfnamefont{P.}~\bibnamefont{Kossacki}},
  \bibinfo{author}{\bibfnamefont{M.}~\bibnamefont{Bertolini}},
  \bibinfo{author}{\bibfnamefont{D.}~\bibnamefont{Ferrand}},
  \bibinfo{author}{\bibfnamefont{J.}~\bibnamefont{Cibert}},
  \bibinfo{author}{\bibfnamefont{S.}~\bibnamefont{Tatarenko}},
  \bibinfo{author}{\bibfnamefont{A.}~\bibnamefont{Wasiela}},
  \bibinfo{author}{\bibfnamefont{J.~A.} \bibnamefont{Gaj}}, \bibnamefont{and}
  \bibinfo{author}{\bibfnamefont{T.}~\bibnamefont{Dietl}},
  \bibinfo{journal}{Phys. Rev. Lett.} \textbf{\bibinfo{volume}{88}},
  \bibinfo{pages}{207204} (\bibinfo{year}{2002}).

\bibitem[{\citenamefont{Liu et~al.}(2004)\citenamefont{Liu, Lim, Titova,
  Wojtowicz, Kutrowski, Yee, Dobrowolska, Furdyna, Potashnik, Stone
  et~al.}}]{Liu04}
\bibinfo{author}{\bibfnamefont{X.}~\bibnamefont{Liu}},
  \bibinfo{author}{\bibfnamefont{W.~L.} \bibnamefont{Lim}},
  \bibinfo{author}{\bibfnamefont{L.~V.} \bibnamefont{Titova}},
  \bibinfo{author}{\bibfnamefont{T.}~\bibnamefont{Wojtowicz}},
  \bibinfo{author}{\bibfnamefont{M.}~\bibnamefont{Kutrowski}},
  \bibinfo{author}{\bibfnamefont{K.~J.} \bibnamefont{Yee}},
  \bibinfo{author}{\bibfnamefont{M.}~\bibnamefont{Dobrowolska}},
  \bibinfo{author}{\bibfnamefont{J.~K.} \bibnamefont{Furdyna}},
  \bibinfo{author}{\bibfnamefont{S.~J.} \bibnamefont{Potashnik}},
  \bibinfo{author}{\bibfnamefont{M.~B.} \bibnamefont{Stone}},
  \bibnamefont{et~al.}, \bibinfo{journal}{Physica E}
  \textbf{\bibinfo{volume}{20}}, \bibinfo{pages}{370} (\bibinfo{year}{2004}).

\bibitem[{\citenamefont{Masmanidis et~al.}(2005)\citenamefont{Masmanidis, Tang,
  Myers, Li, De~Greve, Vermeulen, Van~Roy, and Roukes}}]{Mas05}
\bibinfo{author}{\bibfnamefont{S.~C.} \bibnamefont{Masmanidis}},
  \bibinfo{author}{\bibfnamefont{H.~X.} \bibnamefont{Tang}},
  \bibinfo{author}{\bibfnamefont{E.~B.} \bibnamefont{Myers}},
  \bibinfo{author}{\bibfnamefont{M.}~\bibnamefont{Li}},
  \bibinfo{author}{\bibfnamefont{K.}~\bibnamefont{De~Greve}},
  \bibinfo{author}{\bibfnamefont{G.}~\bibnamefont{Vermeulen}},
  \bibinfo{author}{\bibfnamefont{W.~V.}~\bibnamefont{Roy}}, \bibnamefont{and}
  \bibinfo{author}{\bibfnamefont{M.~L.} \bibnamefont{Roukes}},
  \bibinfo{journal}{Phys. Rev. Lett.} \textbf{\bibinfo{volume}{95}},
  \bibinfo{pages}{187206} (\bibinfo{year}{2005}).

\bibitem[{\citenamefont{Titova et~al.}(2005)\citenamefont{Titova, Kutrowski,
  Liu, Chakarvorty, Lim, Wojtowicz, Furdyna, and Dobrowolska}}]{Tit05}
\bibinfo{author}{\bibfnamefont{L.~V.} \bibnamefont{Titova}},
  \bibinfo{author}{\bibfnamefont{M.}~\bibnamefont{Kutrowski}},
  \bibinfo{author}{\bibfnamefont{X.}~\bibnamefont{Liu}},
  \bibinfo{author}{\bibfnamefont{R.}~\bibnamefont{Chakarvorty}},
  \bibinfo{author}{\bibfnamefont{W.~L.} \bibnamefont{Lim}},
  \bibinfo{author}{\bibfnamefont{T.}~\bibnamefont{Wojtowicz}},
  \bibinfo{author}{\bibfnamefont{J.~K.} \bibnamefont{Furdyna}},
  \bibnamefont{and}
  \bibinfo{author}{\bibfnamefont{M.}~\bibnamefont{Dobrowolska}},
  \bibinfo{journal}{Phys. Rev. B} \textbf{\bibinfo{volume}{72}},
  \bibinfo{pages}{165205} (\bibinfo{year}{2005}).

\bibitem[{\citenamefont{Takamura et~al.}(2002)\citenamefont{Takamura,
  Matsukura, Chiba, and Ohno}}]{Tak02}
\bibinfo{author}{\bibfnamefont{K.}~\bibnamefont{Takamura}},
  \bibinfo{author}{\bibfnamefont{F.}~\bibnamefont{Matsukura}},
  \bibinfo{author}{\bibfnamefont{D.}~\bibnamefont{Chiba}}, \bibnamefont{and}
  \bibinfo{author}{\bibfnamefont{H.}~\bibnamefont{Ohno}},
  \bibinfo{journal}{Appl. Phys. Lett.} \textbf{\bibinfo{volume}{81}},
  \bibinfo{pages}{2590} (\bibinfo{year}{2002}).

\bibitem[{\citenamefont{Daeubler et~al.}(2007)\citenamefont{Daeubler,
  Schwaiger, Glunk, Tabor, Schoch, Sauer, and Limmer}}]{Dae07}
\bibinfo{author}{\bibfnamefont{J.}~\bibnamefont{Daeubler}},
  \bibinfo{author}{\bibfnamefont{S.}~\bibnamefont{Schwaiger}},
  \bibinfo{author}{\bibfnamefont{M.}~\bibnamefont{Glunk}},
  \bibinfo{author}{\bibfnamefont{M.}~\bibnamefont{Tabor}},
  \bibinfo{author}{\bibfnamefont{W.}~\bibnamefont{Schoch}},
  \bibinfo{author}{\bibfnamefont{R.}~\bibnamefont{Sauer}}, \bibnamefont{and}
  \bibinfo{author}{\bibfnamefont{W.}~\bibnamefont{Limmer}},
  \bibinfo{journal}{Physica E, Doi: 10.1016/j.physe.2007.08.049}
  (\bibinfo{year}{2007}).

\bibitem[{\citenamefont{Csontos et~al.}(2005)\citenamefont{Csontos,
  Mih$\acute{\rm a}$ly, Jank$\acute{\rm o}$, Wojtowicz, Liu, and
  Furdyna}}]{Cso05}
\bibinfo{author}{\bibfnamefont{M.}~\bibnamefont{Csontos}},
  \bibinfo{author}{\bibfnamefont{G.}~\bibnamefont{Mih$\acute{\rm a}$ly}},
  \bibinfo{author}{\bibfnamefont{B.}~\bibnamefont{Jank$\acute{\rm o}$}},
  \bibinfo{author}{\bibfnamefont{T.}~\bibnamefont{Wojtowicz}},
  \bibinfo{author}{\bibfnamefont{X.}~\bibnamefont{Liu}}, \bibnamefont{and}
  \bibinfo{author}{\bibfnamefont{J.~K.} \bibnamefont{Furdyna}},
  \bibinfo{journal}{Nature Mater.} \textbf{\bibinfo{volume}{4}},
  \bibinfo{pages}{447} (\bibinfo{year}{2005}).

\bibitem[{\citenamefont{Goennenwein et~al.}(2004)\citenamefont{Goennenwein,
  Wassner, Huebl, Brandt, Philipp, Opel, Gross, Koeder, Schoch, and
  Waag}}]{Goe04}
\bibinfo{author}{\bibfnamefont{S.~T.~B.} \bibnamefont{Goennenwein}},
  \bibinfo{author}{\bibfnamefont{T.~A.} \bibnamefont{Wassner}},
  \bibinfo{author}{\bibfnamefont{H.}~\bibnamefont{Huebl}},
  \bibinfo{author}{\bibfnamefont{M.~S.} \bibnamefont{Brandt}},
  \bibinfo{author}{\bibfnamefont{J.~B.} \bibnamefont{Philipp}},
  \bibinfo{author}{\bibfnamefont{M.}~\bibnamefont{Opel}},
  \bibinfo{author}{\bibfnamefont{R.}~\bibnamefont{Gross}},
  \bibinfo{author}{\bibfnamefont{A.}~\bibnamefont{Koeder}},
  \bibinfo{author}{\bibfnamefont{W.}~\bibnamefont{Schoch}}, \bibnamefont{and}
  \bibinfo{author}{\bibfnamefont{A.}~\bibnamefont{Waag}},
  \bibinfo{journal}{Phys. Rev. Lett.} \textbf{\bibinfo{volume}{92}},
  \bibinfo{pages}{227202} (\bibinfo{year}{2004}).

\bibitem[{\citenamefont{Bouanani-Rahbi
  et~al.}(2003)\citenamefont{Bouanani-Rahbi, Clerjaud, Theys, Lema$\hat{\rm
  i}$tre, and Jomard}}]{Bou03}
\bibinfo{author}{\bibfnamefont{R.}~\bibnamefont{Bouanani-Rahbi}},
  \bibinfo{author}{\bibfnamefont{B.}~\bibnamefont{Clerjaud}},
  \bibinfo{author}{\bibfnamefont{B.}~\bibnamefont{Theys}},
  \bibinfo{author}{\bibfnamefont{A.}~\bibnamefont{Lema$\hat{\rm i}$tre}},
  \bibnamefont{and} \bibinfo{author}{\bibfnamefont{F.}~\bibnamefont{Jomard}},
  \bibinfo{journal}{Physica B} \textbf{\bibinfo{volume}{340-342}},
  \bibinfo{pages}{284} (\bibinfo{year}{2003}).

\bibitem[{\citenamefont{Brandt et~al.}(2004)\citenamefont{Brandt, Goennenwein,
  Wassner, Kohl, Lehner, Huebl, Graf, Stutzmann, Koeder, Schoch
  et~al.}}]{Bra04}
\bibinfo{author}{\bibfnamefont{M.~S.} \bibnamefont{Brandt}},
  \bibinfo{author}{\bibfnamefont{S.~T.~B.} \bibnamefont{Goennenwein}},
  \bibinfo{author}{\bibfnamefont{T.~A.} \bibnamefont{Wassner}},
  \bibinfo{author}{\bibfnamefont{F.}~\bibnamefont{Kohl}},
  \bibinfo{author}{\bibfnamefont{A.}~\bibnamefont{Lehner}},
  \bibinfo{author}{\bibfnamefont{H.}~\bibnamefont{Huebl}},
  \bibinfo{author}{\bibfnamefont{T.}~\bibnamefont{Graf}},
  \bibinfo{author}{\bibfnamefont{M.}~\bibnamefont{Stutzmann}},
  \bibinfo{author}{\bibfnamefont{A.}~\bibnamefont{Koeder}},
  \bibinfo{author}{\bibfnamefont{W.}~\bibnamefont{Schoch}},
  \bibnamefont{et~al.}, \bibinfo{journal}{Appl. Phys. Lett.}
  \textbf{\bibinfo{volume}{84}}, \bibinfo{pages}{2277} (\bibinfo{year}{2004}).

\bibitem[{\citenamefont{Thevenard et~al.}(2005)\citenamefont{Thevenard,
  Largeau, Mauguin, Lema$\hat{\rm i}$tre, and Theys}}]{The05}
\bibinfo{author}{\bibfnamefont{L.}~\bibnamefont{Thevenard}},
  \bibinfo{author}{\bibfnamefont{L.}~\bibnamefont{Largeau}},
  \bibinfo{author}{\bibfnamefont{O.}~\bibnamefont{Mauguin}},
  \bibinfo{author}{\bibfnamefont{A.}~\bibnamefont{Lema$\hat{\rm i}$tre}},
  \bibnamefont{and} \bibinfo{author}{\bibfnamefont{B.}~\bibnamefont{Theys}},
  \bibinfo{journal}{Appl. Phys. Lett.} \textbf{\bibinfo{volume}{87}},
  \bibinfo{pages}{182506} (\bibinfo{year}{2005}).

\bibitem[{\citenamefont{Farshchi et~al.}(2008)\citenamefont{Farshchi, Dubon,
  Hwang, Misra, Grigoropoulos, and Ashby}}]{Far08}
\bibinfo{author}{\bibfnamefont{R.}~\bibnamefont{Farshchi}},
  \bibinfo{author}{\bibfnamefont{O.~D.} \bibnamefont{Dubon}},
  \bibinfo{author}{\bibfnamefont{D.~J.} \bibnamefont{Hwang}},
  \bibinfo{author}{\bibfnamefont{N.}~\bibnamefont{Misra}},
  \bibinfo{author}{\bibfnamefont{C.~P.} \bibnamefont{Grigoropoulos}},
  \bibnamefont{and} \bibinfo{author}{\bibfnamefont{P.~D.} \bibnamefont{Ashby}},
  \bibinfo{journal}{Appl. Phys. Lett.} \textbf{\bibinfo{volume}{92}},
  \bibinfo{pages}{012517} (\bibinfo{year}{2008}).

\bibitem[{\citenamefont{Wenisch et~al.}(2007)\citenamefont{Wenisch, Gould,
  Ebel, Storz, Pappert, Schmidt, Kumpf, Schmidt, Brunner, and
  Molenkamp}}]{Wen07}
\bibinfo{author}{\bibfnamefont{J.}~\bibnamefont{Wenisch}},
  \bibinfo{author}{\bibfnamefont{C.}~\bibnamefont{Gould}},
  \bibinfo{author}{\bibfnamefont{L.}~\bibnamefont{Ebel}},
  \bibinfo{author}{\bibfnamefont{J.}~\bibnamefont{Storz}},
  \bibinfo{author}{\bibfnamefont{K.}~\bibnamefont{Pappert}},
  \bibinfo{author}{\bibfnamefont{M.~J.} \bibnamefont{Schmidt}},
  \bibinfo{author}{\bibfnamefont{C.}~\bibnamefont{Kumpf}},
  \bibinfo{author}{\bibfnamefont{G.}~\bibnamefont{Schmidt}},
  \bibinfo{author}{\bibfnamefont{K.}~\bibnamefont{Brunner}}, \bibnamefont{and}
  \bibinfo{author}{\bibfnamefont{L.~W.} \bibnamefont{Molenkamp}},
  \bibinfo{journal}{Phys. Rev. Lett.} \textbf{\bibinfo{volume}{99}},
  \bibinfo{pages}{077201} (\bibinfo{year}{2007}).

\bibitem[{\citenamefont{Botters et~al.}(2006)\citenamefont{Botters, Giesen,
  Podbielski, Bach, Schmidt, Molenkamp, and Grundler}}]{Bot06}
\bibinfo{author}{\bibfnamefont{B.}~\bibnamefont{Botters}},
  \bibinfo{author}{\bibfnamefont{F.}~\bibnamefont{Giesen}},
  \bibinfo{author}{\bibfnamefont{J.}~\bibnamefont{Podbielski}},
  \bibinfo{author}{\bibfnamefont{P.}~\bibnamefont{Bach}},
  \bibinfo{author}{\bibfnamefont{G.}~\bibnamefont{Schmidt}},
  \bibinfo{author}{\bibfnamefont{L.~W.} \bibnamefont{Molenkamp}},
  \bibnamefont{and} \bibinfo{author}{\bibfnamefont{D.}~\bibnamefont{Grundler}},
  \bibinfo{journal}{Appl. Phys. Lett.} \textbf{\bibinfo{volume}{89}},
  \bibinfo{pages}{242505} (\bibinfo{year}{2006}).

\bibitem[{\citenamefont{Brandlmaier et~al.}(2008)\citenamefont{Brandlmaier,
  Gepr\"{a}gs, Weiler, Boger, Opel, Huebl, Bihler, Brandt, Botters, Grundler
  et~al.}}]{Bra08}
\bibinfo{author}{\bibfnamefont{A.}~\bibnamefont{Brandlmaier}},
  \bibinfo{author}{\bibfnamefont{S.}~\bibnamefont{Gepr\"{a}gs}},
  \bibinfo{author}{\bibfnamefont{M.}~\bibnamefont{Weiler}},
  \bibinfo{author}{\bibfnamefont{A.}~\bibnamefont{Boger}},
  \bibinfo{author}{\bibfnamefont{M.}~\bibnamefont{Opel}},
  \bibinfo{author}{\bibfnamefont{H.}~\bibnamefont{Huebl}},
  \bibinfo{author}{\bibfnamefont{C.}~\bibnamefont{Bihler}},
  \bibinfo{author}{\bibfnamefont{M.~S.} \bibnamefont{Brandt}},
  \bibinfo{author}{\bibfnamefont{B.}~\bibnamefont{Botters}},
  \bibinfo{author}{\bibfnamefont{D.}~\bibnamefont{Grundler}},
  \bibnamefont{et~al.}, \bibinfo{journal}{Phys. Rev. B}
  \textbf{\bibinfo{volume}{77}}, \bibinfo{pages}{104445}
  (\bibinfo{year}{2008}).

\bibitem[{\citenamefont{Goennenwein et~al.}(2008)\citenamefont{Goennenwein,
  Althammer, Bihler, Brandlmaier, Gepr\"{a}gs, Opel, Schoch, Limmer, Gross, and
  Brandt}}]{Goe08}
\bibinfo{author}{\bibfnamefont{S.~T.~B.} \bibnamefont{Goennenwein}},
  \bibinfo{author}{\bibfnamefont{M.}~\bibnamefont{Althammer}},
  \bibinfo{author}{\bibfnamefont{C.}~\bibnamefont{Bihler}},
  \bibinfo{author}{\bibfnamefont{A.}~\bibnamefont{Brandlmaier}},
  \bibinfo{author}{\bibfnamefont{S.}~\bibnamefont{Gepr\"{a}gs}},
  \bibinfo{author}{\bibfnamefont{M.}~\bibnamefont{Opel}},
  \bibinfo{author}{\bibfnamefont{W.}~\bibnamefont{Schoch}},
  \bibinfo{author}{\bibfnamefont{W.}~\bibnamefont{Limmer}},
  \bibinfo{author}{\bibfnamefont{R.}~\bibnamefont{Gross}}, \bibnamefont{and}
  \bibinfo{author}{\bibfnamefont{M.~S.} \bibnamefont{Brandt}},
  \bibinfo{journal}{phys. stat. sol. (RRL)} \textbf{\bibinfo{volume}{2}},
  \bibinfo{pages}{96} (\bibinfo{year}{2008}).

\bibitem[{\citenamefont{Rushforth et~al.}(2008)\citenamefont{Rushforth,
  De~Ranieri, Zemen, Wunderlich, Edmonds, King, Ahmad, Campion, Foxon,
  Gallagher et~al.}}]{Rus08}
\bibinfo{author}{\bibfnamefont{A.~W.} \bibnamefont{Rushforth}},
  \bibinfo{author}{\bibfnamefont{E.}~\bibnamefont{De~Ranieri}},
  \bibinfo{author}{\bibfnamefont{J.}~\bibnamefont{Zemen}},
  \bibinfo{author}{\bibfnamefont{J.}~\bibnamefont{Wunderlich}},
  \bibinfo{author}{\bibfnamefont{K.~W.} \bibnamefont{Edmonds}},
  \bibinfo{author}{\bibfnamefont{C.~S.} \bibnamefont{King}},
  \bibinfo{author}{\bibfnamefont{E.}~\bibnamefont{Ahmad}},
  \bibinfo{author}{\bibfnamefont{R.~P.} \bibnamefont{Campion}},
  \bibinfo{author}{\bibfnamefont{C.~T.} \bibnamefont{Foxon}},
  \bibinfo{author}{\bibfnamefont{B.~L.} \bibnamefont{Gallagher}},
  \bibnamefont{et~al.}, \bibinfo{journal}{arXiv:0801.0886v2}
  (\bibinfo{year}{2008}).

\bibitem[{\citenamefont{Overby et~al.}(2008)\citenamefont{Overby, Chernyshov,
  Rokhinson, Liu, and Furdyna}}]{Ove08}
\bibinfo{author}{\bibfnamefont{M.}~\bibnamefont{Overby}},
  \bibinfo{author}{\bibfnamefont{A.}~\bibnamefont{Chernyshov}},
  \bibinfo{author}{\bibfnamefont{L.~P.} \bibnamefont{Rokhinson}},
  \bibinfo{author}{\bibfnamefont{X.}~\bibnamefont{Liu}}, \bibnamefont{and}
  \bibinfo{author}{\bibfnamefont{J.~K.} \bibnamefont{Furdyna}},
  \bibinfo{journal}{arXiv:0801.4191v1}  (\bibinfo{year}{2008}).

\bibitem[{\citenamefont{Limmer et~al.}(2006)\citenamefont{Limmer, Glunk,
  Daeubler, Hummel, Schoch, Sauer, Bihler, Huebl, Brandt, and
  Goennenwein}}]{Lim06}
\bibinfo{author}{\bibfnamefont{W.}~\bibnamefont{Limmer}},
  \bibinfo{author}{\bibfnamefont{M.}~\bibnamefont{Glunk}},
  \bibinfo{author}{\bibfnamefont{J.}~\bibnamefont{Daeubler}},
  \bibinfo{author}{\bibfnamefont{T.}~\bibnamefont{Hummel}},
  \bibinfo{author}{\bibfnamefont{W.}~\bibnamefont{Schoch}},
  \bibinfo{author}{\bibfnamefont{R.}~\bibnamefont{Sauer}},
  \bibinfo{author}{\bibfnamefont{C.}~\bibnamefont{Bihler}},
  \bibinfo{author}{\bibfnamefont{H.}~\bibnamefont{Huebl}},
  \bibinfo{author}{\bibfnamefont{M.~S.} \bibnamefont{Brandt}},
  \bibnamefont{and} \bibinfo{author}{\bibfnamefont{S.~T.~B.}
  \bibnamefont{Goennenwein}}, \bibinfo{journal}{Phys. Rev. B}
  \textbf{\bibinfo{volume}{74}}, \bibinfo{pages}{205205}
  (\bibinfo{year}{2006}).

\bibitem[{\citenamefont{Bihler et~al.}(2006)\citenamefont{Bihler, Huebl,
  Brandt, Goennenwein, Reinwald, Wurstbauer, D\"{o}ppe, Weiss, and
  Wegscheider}}]{Bihler3112006}
\bibinfo{author}{\bibfnamefont{C.}~\bibnamefont{Bihler}},
  \bibinfo{author}{\bibfnamefont{H.}~\bibnamefont{Huebl}},
  \bibinfo{author}{\bibfnamefont{M.~S.} \bibnamefont{Brandt}},
  \bibinfo{author}{\bibfnamefont{S.~T.~B.} \bibnamefont{Goennenwein}},
  \bibinfo{author}{\bibfnamefont{M.}~\bibnamefont{Reinwald}},
  \bibinfo{author}{\bibfnamefont{U.}~\bibnamefont{Wurstbauer}},
  \bibinfo{author}{\bibfnamefont{M.}~\bibnamefont{D\"{o}ppe}},
  \bibinfo{author}{\bibfnamefont{D.}~\bibnamefont{Weiss}}, \bibnamefont{and}
  \bibinfo{author}{\bibfnamefont{W.}~\bibnamefont{Wegscheider}},
  \bibinfo{journal}{Appl. Phys. Lett.} \textbf{\bibinfo{volume}{89}},
  \bibinfo{pages}{012507} (\bibinfo{year}{2006}).

\bibitem[{\citenamefont{Bihler et~al.}(2007)\citenamefont{Bihler, Kraus, Huebl,
  Brandt, Goennenwein, Opel, Scarpulla, Stone, Farshchi, and Dubon}}]{Bih07}
\bibinfo{author}{\bibfnamefont{C.}~\bibnamefont{Bihler}},
  \bibinfo{author}{\bibfnamefont{M.}~\bibnamefont{Kraus}},
  \bibinfo{author}{\bibfnamefont{H.}~\bibnamefont{Huebl}},
  \bibinfo{author}{\bibfnamefont{M.~S.} \bibnamefont{Brandt}},
  \bibinfo{author}{\bibfnamefont{S.~T.~B.} \bibnamefont{Goennenwein}},
  \bibinfo{author}{\bibfnamefont{M.}~\bibnamefont{Opel}},
  \bibinfo{author}{\bibfnamefont{M.~A.} \bibnamefont{Scarpulla}},
  \bibinfo{author}{\bibfnamefont{P.~R.} \bibnamefont{Stone}},
  \bibinfo{author}{\bibfnamefont{R.}~\bibnamefont{Farshchi}}, \bibnamefont{and}
  \bibinfo{author}{\bibfnamefont{O.~D.} \bibnamefont{Dubon}},
  \bibinfo{journal}{Phys. Rev. B} \textbf{\bibinfo{volume}{75}},
  \bibinfo{pages}{214419} (\bibinfo{year}{2007}).

\bibitem[{\citenamefont{Yamada et~al.}(2006)\citenamefont{Yamada, Chiba,
  Matsukura, Yakata, and Ohno}}]{Yamada06}
\bibinfo{author}{\bibfnamefont{T.}~\bibnamefont{Yamada}},
  \bibinfo{author}{\bibfnamefont{D.}~\bibnamefont{Chiba}},
  \bibinfo{author}{\bibfnamefont{F.}~\bibnamefont{Matsukura}},
  \bibinfo{author}{\bibfnamefont{S.}~\bibnamefont{Yakata}}, \bibnamefont{and}
  \bibinfo{author}{\bibfnamefont{H.}~\bibnamefont{Ohno}},
  \bibinfo{journal}{phys. stat. sol. (c)} \textbf{\bibinfo{volume}{3}},
  \bibinfo{pages}{4086} (\bibinfo{year}{2006}).

\bibitem[{\citenamefont{Goennenwein et~al.}(2005)\citenamefont{Goennenwein,
  Russo, Morpurgo, Klapwijk, Van~Roy, and De~Boeck}}]{Goe05}
\bibinfo{author}{\bibfnamefont{S.~T.~B.} \bibnamefont{Goennenwein}},
  \bibinfo{author}{\bibfnamefont{S.}~\bibnamefont{Russo}},
  \bibinfo{author}{\bibfnamefont{A.~F.} \bibnamefont{Morpurgo}},
  \bibinfo{author}{\bibfnamefont{T.~M.} \bibnamefont{Klapwijk}},
  \bibinfo{author}{\bibfnamefont{W.}~\bibnamefont{Van~Roy}}, \bibnamefont{and}
  \bibinfo{author}{\bibfnamefont{J.}~\bibnamefont{De~Boeck}},
  \bibinfo{journal}{Phys. Rev. B} \textbf{\bibinfo{volume}{71}},
  \bibinfo{pages}{193306} (\bibinfo{year}{2005}).

\bibitem[{\citenamefont{Wu et~al.}(2008)\citenamefont{Wu, Wei,
  Johnston-Halperin, Awschalom, and Shi}}]{Wu08}
\bibinfo{author}{\bibfnamefont{D.}~\bibnamefont{Wu}},
  \bibinfo{author}{\bibfnamefont{P.}~\bibnamefont{Wei}},
  \bibinfo{author}{\bibfnamefont{E.}~\bibnamefont{Johnston-Halperin}},
  \bibinfo{author}{\bibfnamefont{D.~D.} \bibnamefont{Awschalom}},
  \bibnamefont{and} \bibinfo{author}{\bibfnamefont{J.}~\bibnamefont{Shi}},
  \bibinfo{journal}{Phys. Rev. B} \textbf{\bibinfo{volume}{77}},
  \bibinfo{pages}{125320} (\bibinfo{year}{2008}).

\bibitem[{\citenamefont{Blakemore}(1982)}]{Bla82}
\bibinfo{author}{\bibfnamefont{J.~S.} \bibnamefont{Blakemore}},
  \bibinfo{journal}{J. Appl. Phys.} \textbf{\bibinfo{volume}{53}},
  \bibinfo{pages}{R123} (\bibinfo{year}{1982}).

\bibitem[{\citenamefont{Matsuoka et~al.}(1994)\citenamefont{Matsuoka, Hoshino,
  and Ono}}]{Mat94}
\bibinfo{author}{\bibfnamefont{M.}~\bibnamefont{Matsuoka}},
  \bibinfo{author}{\bibfnamefont{K.}~\bibnamefont{Hoshino}}, \bibnamefont{and}
  \bibinfo{author}{\bibfnamefont{K.}~\bibnamefont{Ono}}, \bibinfo{journal}{J.
  Appl. Phys.} \textbf{\bibinfo{volume}{76}}, \bibinfo{pages}{1768}
  (\bibinfo{year}{1994}).

\bibitem[{\citenamefont{Alexe and Hesse}(2006)}]{Ale06}
\bibinfo{author}{\bibfnamefont{M.}~\bibnamefont{Alexe}} \bibnamefont{and}
  \bibinfo{author}{\bibfnamefont{D.}~\bibnamefont{Hesse}}, \bibinfo{journal}{J.
  Mater. Sci.} \textbf{\bibinfo{volume}{41}}, \bibinfo{pages}{1}
  (\bibinfo{year}{2006}).

\bibitem[{\citenamefont{Morita et~al.}(2004)\citenamefont{Morita, Wagatsuma,
  Cho, Morioka, Funakubo, and Setter}}]{Mor04}
\bibinfo{author}{\bibfnamefont{T.}~\bibnamefont{Morita}},
  \bibinfo{author}{\bibfnamefont{Y.}~\bibnamefont{Wagatsuma}},
  \bibinfo{author}{\bibfnamefont{Y.}~\bibnamefont{Cho}},
  \bibinfo{author}{\bibfnamefont{H.}~\bibnamefont{Morioka}},
  \bibinfo{author}{\bibfnamefont{H.}~\bibnamefont{Funakubo}}, \bibnamefont{and}
  \bibinfo{author}{\bibfnamefont{N.}~\bibnamefont{Setter}},
  \bibinfo{journal}{Appl. Phys. Lett.} \textbf{\bibinfo{volume}{84}},
  \bibinfo{pages}{5094} (\bibinfo{year}{2004}).

\bibitem[{\citenamefont{Morita and Cho}(2006)}]{Mor06}
\bibinfo{author}{\bibfnamefont{T.}~\bibnamefont{Morita}} \bibnamefont{and}
  \bibinfo{author}{\bibfnamefont{Y.}~\bibnamefont{Cho}},
  \bibinfo{journal}{Appl. Phys. Lett.} \textbf{\bibinfo{volume}{88}},
  \bibinfo{pages}{112908} (\bibinfo{year}{2006}).

\bibitem[{\citenamefont{Furukawa}(1989)}]{Fur89}
\bibinfo{author}{\bibfnamefont{T.}~\bibnamefont{Furukawa}},
  \bibinfo{journal}{Phase Transition} \textbf{\bibinfo{volume}{18}}
  (\bibinfo{year}{1989}).

\bibitem[{\citenamefont{Naber et~al.}(2005)\citenamefont{Naber, Tanase, Blom,
  Gelinck, Marsman, Touslager, Setayesh, and De~Leeuw}}]{Nab05}
\bibinfo{author}{\bibfnamefont{R.~C.~G.} \bibnamefont{Naber}},
  \bibinfo{author}{\bibfnamefont{C.}~\bibnamefont{Tanase}},
  \bibinfo{author}{\bibfnamefont{P.~W.~M.} \bibnamefont{Blom}},
  \bibinfo{author}{\bibfnamefont{G.~H.} \bibnamefont{Gelinck}},
  \bibinfo{author}{\bibfnamefont{A.~W.} \bibnamefont{Marsman}},
  \bibinfo{author}{\bibfnamefont{F.~J.} \bibnamefont{Touslager}},
  \bibinfo{author}{\bibfnamefont{S.}~\bibnamefont{Setayesh}}, \bibnamefont{and}
  \bibinfo{author}{\bibfnamefont{D.~M.} \bibnamefont{De~Leeuw}},
  \bibinfo{journal}{Nature Mater.} \textbf{\bibinfo{volume}{4}},
  \bibinfo{pages}{243} (\bibinfo{year}{2005}).

\bibitem[{\citenamefont{Chikazumi}(1964)}]{Chi64}
\bibinfo{author}{\bibfnamefont{S.}~\bibnamefont{Chikazumi}},
  \emph{\bibinfo{title}{Physics of Magnetism}} (\bibinfo{publisher}{John Wiley
  \& Sons, New York}, \bibinfo{year}{1964}).

\end{thebibliography}
\end{document}